\documentclass[aps, prd, showpacs, superscriptaddress, preprintnumbers, nofootinbib, twocolumn]{revtex4-2}
\usepackage[pdftex]{graphicx, color}
\usepackage{amsmath, amssymb, a
mscd, latexsym, bm, braket,times}
\usepackage[colorlinks=true,pdfstartview=FitV,linkcolor=blue,citecolor=magenta,urlcolor=blue,bookmarks=true,bookmarksnumbered=true]{hyperref}

\begin{document}

\title{Hydrodynamic attractor in a Hubble expansion}
\author{Zhiwei Du}
\address{Department of Physics and Center for Field Theory and Particle Physics, Fudan University, Shanghai, 200433, China }
\author{Xu-Guang Huang}
\email{huangxuguang@fudan.edu.cn}
\address{Department of Physics and Center for Field Theory and Particle Physics, Fudan University, Shanghai, 200433, China }
\address{Key Laboratory of Nuclear Physics and Ion-beam Application (MOE), Fudan University, Shanghai 200433, China }
\author{Hidetoshi Taya}
\email{hidetoshi.taya@riken.jp}
\address{iTHEMS, RIKEN, 2-1 Hirosawa, Wako, Saitama 351-0198, Japan}
\address{Research and Education Center for Natural Sciences, Keio University 4-1-1 Hiyoshi, Kohoku-ku, Yokohama, Kanagawa 223-8521, Japan}

\date{\today}

\begin{abstract}
We analytically investigate hydrodynamic attractor solutions in both M\"{u}ller-Israel-Stewart (MIS) and kinetic theories in a viscous fluid system undergoing a Hubble expansion with a fixed expansion rate.  We show that the gradient expansion for the MIS theory and the Chapman-Enskog expansion for the Boltzmann equation within the relaxation time approximation are factorially divergent and obtain hydrodynamic attractor solutions by applying the Borel resummation technique to those asymptotic divergent series.  In both theories, we find that the hydrodynamic attractor solutions are globally attractive and only the first-order non-hydrodynamic mode exists.  We also find that the hydrodynamic attractor solutions in the two theories disagree with each other when gradients become large, and that the speed of the attraction is different.  Similarities and differences from hydrodynamic attractors in the Bjorken and Gubser flows are also discussed.  Our results push the idea of far-from-equilibrium hydrodynamics in systems undergoing a Hubble expansion.
\end{abstract}

\maketitle

\section{Introduction}
Hydrodynamics provides a fundamental framework to describe fluid dynamics and has a wide range of applications across many branches of physics.  For example, in cosmology, hydrodynamics within the ideal approximation has been widely used to model the matter contents of the universe~\cite{Weinberg:1972kfs}.  Integration of viscosities into hydrodynamics has been argued to be important for understanding, e.g., the inflationary phase of the early universe~\cite{Brevik_2017}.  In astrophysics, relativistic hydrodynamics and magnetohydrodynamics are indispensable tools to simulate the dynamics of black-hole accretion, the explosion of supernovas, and the formation of compact stars~\cite{Rezzolla}.  From the beginning days of heavy-ion collision physics, theorists have been using hydrodynamic models to predict, describe, and simulate the expansion of the quark-gluon plasma (QGP) created in the collisions~\cite{Heinz:2013th,Gale:2013da,Romatschke:2017ejr, Florkowski_2018}.  Theoretical predictions for such as radial flow and elliptic flow have gained supporting evidences from experimental data; see Ref.~\cite{Shen2020} for a recent review.

Hydrodynamics is an effective theory for infrared modes (hydrodynamic modes) that survive in the long wavelength and low-frequency limit.  In the standard formulation, hydrodynamics is defined as a low-order truncation of the gradient expansion near equilibrium.  For example, the well-known Navier-Stokes (NS) theory, yielded by the truncation at the first order, captures effects of shear and bulk viscosities.  However, the relativistic version of the NS theory contains acausal modes, which lead to numerical instability in practical use~\cite{Hiscock:1983zz,Pu:2009fj}.  A minimal extension of the NS theory that preserves relativistic causality is the M\"{u}ller-Israel-Stewart (MIS) theory~\cite{Muller:1967zza,ISRAEL1979341}, which we shall study in Sec.~\ref{sec:3}.  Given that hydrodynamics is based on the gradient expansion procedure, one may expect that it applies only when higher-order gradients are smaller so that the expansion series is convergent and that hydrodynamics is inapplicable in far-from-equilibrium situations where gradients are large.  On the contrary to those naive expectations, substantial evidences have been accumulated that the applicability of hydrodynamics is not necessarily limited to systems with small gradients~\cite{Florkowski_2018}.  In particular, the recent heavy-ion collision experiments have witnessed ``unreasonable" effectiveness of hydrodynamics in small systems such as pp and pA collisions, where large gradients arise~\cite{Weller:2017tsr,Mantysaari:2017cni,Habich_2016,Nagle:2013lja}.  Those findings motivated people to re-consider the applicability of hydrodynamics.  It turned out that the gradient expansion can be factorially divergent, implying that the naive formulation of hydrodynamics has zero radius of convergence and is ill-defined~\cite{Heller_2013, Heller_2015}.  Such a divergence, however, does not ruin the notion of hydrodynamics, but suggests that one needs to systematically resum higher-order gradient contributions to re-formulate hydrodynamics in a well-defined manner.

The Borel-resummation technique, developed by \'{E}calle in mathematics in 1980's~\cite{Ecalle}, is one of the possible methods to carry out the resummation.  This mathematical technique has been successfully applied to a variety of physics problems~\cite{ANICETO20191}, including  hydrodynamics~\cite{Heller_2015,Aniceto2016, Ba_ar_2015, PhysRevD.97.091503,denicol2017analytical, romatschke_2,Blaizot:2017lht,Behtash18, Denicol_2019}, quantum mechanics \cite{Zinn-Justin:1981:PsloqmaftApr,Jentschura_2004,Misumi_2015}, and quantum field theory \cite{Dunne_2012,AnicSchi15:NonpeAmbigandRealiResurTrans,Pasquetti_2010,Taya:2020dco}.  In hydrodynamics, it was found that the Borel resummation systematically includes not only higher-order gradients near equilibrium but also non-perturbative and off-equilibrium effects, coming from the so-called non-hydrodynamic modes~\cite{Aniceto2016, Ba_ar_2015, PhysRevD.97.091503, Florkowski_2018}.  The non-hydrodynamic modes damp out in a characteristic time/length-scale set by the so-called instanton action in the language of the Borel resummation.  In the infrared limit, only an attractor solution, which is called hydrodynamic attractor, survives and asymptotes the standard hydrodynamics obtained as a low-order truncation of the gradient expansion.  The hydrodynamic attractor can capture essential features of the dynamical evolution of systems even with large gradients and thus provides a possible explanation to the ``unreasonable" effectiveness of hydrodynamics observed in heavy-ion collisions.

The Bjorken and Gubser flows, which are commonly used to describe the anisotropic expansion of QGP in heavy-ion collisions, are popular systems for the study of hydrodynamic attractors~\cite{Heller_2015,Ba_ar_2015,denicol2017analytical,romatschke_2,Blaizot:2017lht,Behtash18, Denicol_2019,Romatschke:2017acs,Mitra:2020mei,Blaizot:2020gql,Behtash:2019qtk,Jaiswal:2019cju,Kurkela:2019set,Blaizot:2019scw,Behtash:2019txb,Strickland:2018ayk,Blaizot:2017ucy,Dash:2020zqx}.  Properties of hydrodynamic attractors change depending on the flow profile.  For example, the Bjorken and Gubser flows have different basins of attraction due to the difference in the dimension of the phase space \cite{Behtash18}.  To deepen our understanding of hydrodynamic attractors, it is, therefore, important to understand possible attractors in other flow systems.  Such a direction may also be important to push the idea of far-from-equilibrium hydrodynamics not only in the QGP fluid created in heavy-ion collisions, but also in other fluid systems appearing in various physical problems.

In this paper, we will, for the first time, study possible attractor solutions in a system undergoing Hubble expansion, a standard model for isotropic expansion of the universe in cosmology~\cite{Weinberg:1972kfs}.  The hydrodynamic theory of Hubble expansion may be applicable for the cosmic quantum chromodynamics (QCD) fluid in the early universe~\cite{Aoki_2006}. Shortly after the Big Bang, the primordial QGP at very high temperature and in a far-from-equilibrium state may arise.  The authors of Ref.~\cite{Buchel:2016cbj} studied the entropy production in the primordial QGP based on holographic theory and found that the simple hydrodynamic gradient expansion cannot describe the entropy production, meaning that non-hydrodynamic modes are necessary to be included via the resurgence procedure.  In this paper, we will derive explicitly the hydrodynamic attractors and non-hydrodynamic modes of the viscous Hubble fluid system from both MIS and kinetic theories.

While hydrodynamics describes the system macroscopically, the kinetic theory focuses on the microscopic process and provides a complete description of drifting, collision, and streaming processes of particles in the fluid.  With such an advantage, the Boltzmann equation is often used to describe far-from-equilibrium evolution of fluid systems.  In the context of Hubble flow, with the method of moments~\cite{grad1963}, the exact solution to relativistic Boltzmann equation was obtained~\cite{Bazow:2016:ASBEES} (with the covariant treatment) and the nonlinear dynamics of a massless fluid is studied with the method of moments~\cite{Bazow:2016:NdrBeFs}.  In this paper, instead of using the method of moments, we will adopt the Chapman-Enskog (CE) method~\cite{chapman53,CercignaniKremer} to solve the Boltzmann equation (with the contravariant treatment) for a Hubble expansion and show that the CE expansion is divergent but can be resummed via the resurgence procedure during which the hydrodynamic attractor emerges.

The paper is organized as follows.  In Sec.~\ref{sec:2}, we give a brief introduction to the Borel resummation technique.  In Sec.~\ref{sec:3} and Sec.~\ref{sec:4}, we consider a viscous fluid system under a Hubble expansion and explore its hydrodynamic attractor solutions with the MIS and kinetic theories, respectively.  Our discussion and summary is given in Sec.~\ref{sec:5}.

\section{Reminder: Borel resummation} \label{sec:2}

In this section, we provide a brief reminder to the Borel resummation technique, which we will use in the following sections to obtain the hydrodynamic attractor solutions in a Hubble-expanding system. For a detailed introduction to the Borel resummation technique and the mathematical theory of resurgence, we recommend Refs.~\cite{Aniceto:2011nu, AnicSchi15:NonpeAmbigandRealiResurTrans}.

Suppose we have a physical observable $F \in {\mathbb R}$ and consider its formal perturbative expansion with respect to some small real positive parameter $0<\epsilon \ll 1$ as
\begin{align}
	F(\epsilon) \simeq  \sum_{k=0}^{\infty} F^{(0)}_k \epsilon^{k} =: F^{(0)}(\epsilon) \ , \label{eq1}
\end{align}
with real-valued series coefficients $F^{(0)}_k$'s~\footnote{For simplicity, we assume that the series expansion (\ref{eq1}) does not start with negative or fractional powers of $\epsilon$ because such a case is not relevant in our problem as we will see later.}. Note that we used ``$\simeq$," instead of an equality, to denote that the expansion (\ref{eq1}) is formal.
When the series coefficients are factorially divergent at large $k$, $F_k^{(0)} \sim k!$ (i.e., Gevrey-1 type divergence~\cite{ChermanKoroteevUnsal}), the perturbative expansion $F^{(0)}$ is ill-defined, since it has zero radius of convergence.  The Borel resummation is a specific technique to cure this problem, which relates the ill-defined perturbative expansion $F^{(0)}$ to some well-defined function that includes non-perturbative information.

To carry out the Borel resummation, we introduce the Borel transform of the perturbative expansion $F^{(0)}$ as
\begin{align}
	{\mathcal B}[F^{(0)}](s) := \sum_{k=0}^{\infty} \frac{F^{(0)}_k}{\Gamma(k+1)} s^{k}\ . \label{eq2}
\end{align}
The Borel transform ${\mathcal B}[F^{(0)}]$ is convergent around $s \sim 0$ and can be analytically continued to the complex Borel plane ($s$-plane) from $s=0$ until it meets singularities such as poles and branch cuts.  It is typical that the Borel transform ${\mathcal B}[F^{(0)}]$ has singularities on the real axis, which contain non-perturbative information of the original observable $F$ as we will see later.

The Borel resummation is defined as the Laplace transform of the Borel transform ${\mathcal B}[F^{(0)}]$,
\begin{align}
	\mathcal{S} [F^{(0)}](\epsilon)
		:= \int_0^{\infty} \frac{ds}{\epsilon} e^{-s/\epsilon} {\mathcal B}[F^{(0)}](s)\ , \label{eq4}
\end{align}
where we assume for the moment that the Borel transform ${\mathcal B}[F^{(0)}]$ does not have any singularities on the positive real axis.  In such a case, the formal perturbative series $F^{(0)}$ is said to be {\it Borel summable}, and the Borel resummation (\ref{eq4}) gives a unique and well-defined function whose asymptotic expansion around $\epsilon \sim 0$ is given by $F^{(0)}$.  Note that the Borel resummation recovers the original series when $F^{(0)}$ is a well-defined series having a finite radius of convergence [for which case, the Laplace transform can be regarded as an inverse operation of the Borel transform (\ref{eq2})].

Typically, the Borel transform $\mathcal{B}[F^{(0)}](s)$ has singularities on the positive real axis, for which $F^{(0)}$ is non-Borel summable along the direction ${\rm arg}\,s=0$.  These singularities render the naive Borel resummation (\ref{eq4}) ill-defined.  To make it well-defined, one has to modify the integration contour to avoid the singularities.  We denote the contour turning above (below) the real axis as the $+$($-$)-contour, and correspondingly the Laplace transform as,
\begin{align}
	\mathcal{S}_\pm [F^{(0)}](\epsilon)
		:= \int_0^{\infty \pm i0^+} \frac{ds}{\epsilon} e^{-s/\epsilon} {\mathcal B}[F^{(0)}](s) \ . \label{eq5}
\end{align}
The two Borel resummations ${\mathcal S}_\pm[F^{(0)}]$ (sometimes called {\it lateral Borel resummations}) yield different results depending on the choice of $+$- and $-$-contours due to  contributions from the singularities.  The difference is purely imaginary and is non-perturbative in terms of the expansion parameter $\epsilon$.  For instance, if the Borel transform $\mathcal{B}[F^{(0)}]$ has a simple pole on the positive real axis $s=s_0>0$, its contribution to $\mathcal{S}_\pm [F^{(0)}](\epsilon)$ reads $\mp i\pi e^{-s_0/\epsilon}{\rm Res}\,\mathcal{B}[F^{(0)}](s_0)/\epsilon \in i {\mathbb R}$.  The imaginary difference leads to problematic ambiguity for ``predicting" $F$ because we neither have a preferential choice between the integration contours $\pm$ nor expect any imaginary result for $F$, which is real-valued.  We note that it is natural that the imaginary ambiguity is non-perturbative because the original perturbative expansion $F^{(0)}$ does not include any non-perturbative terms and so it can be inaccurate for them.  This observation implies that we need to systematically add non-perturbative terms to the perturbative expansion $F^{(0)}$ to resolve the imaginary ambiguity.  Trans-series ansatz introduced below is a convenient way to achieve this aim.

A trans-series $\tilde{F}$ is an augmented perturbative series with non-perturbative factors $e^{-n A/\epsilon}$ included.  Its general form  \footnote{In general, a trans-series can include logarithmic factors as well \cite{dorigoni14}.  For our purpose, such factors are not essential, so we omit them for simplicity.} is
\begin{align}
	F(\epsilon)
		\simeq \sum_{n=0}^{\infty} \frac{\sigma^n}{\epsilon^n} e^{-n A/\epsilon} F^{(n)}(\epsilon)
		=: \tilde{F}(\epsilon; A, \sigma) \ . \label{eq6}
\end{align}
The quantities $A \in {\mathbb R}$ is the so-called instanton action, named after an analogy with the instanton calculus in quantum mechanics \cite{Dunne:2014bca}, and $\sigma \in {\mathbb C}$ is a trans-series parameter.  $A$ and $\sigma$ can be fixed uniquely by requiring that the lateral Borel resummations (\ref{eq5}) are real-valued and free from the imaginary ambiguity and by initial/boundary conditions for the observable $F$, as we explain below.  $F^{(n)}$'s are formal perturbative expansions on top of the non-trivial instanton backgrounds $e^{-nA/\epsilon}$ and are parametrized in the same manner as $F^{(0)}$ as
\begin{align}
	F^{(n)}(\epsilon) :=  \sum_{k=0}^{\infty} F^{(n)}_k \epsilon^k \ . \label{eq7}
\end{align}

With the trans-series ansatz (\ref{eq6}), one can systematically cancel the aforementioned imaginary ambiguity by successively taking into account the instanton contributions and construct a resummation scheme to obtain an unambiguous real-valued function for the original physical observable $F$.  We begin with taking into account the one-instanton sector only and compute its lateral Borel resummations,
\begin{align}
	&\mathcal{S}_\pm[F^{(0)}+\sigma \epsilon^{-1} e^{-A/\epsilon }F^{(1)}] \nonumber\\
		&= {\rm Re}\, \mathcal{S}_{\pm}[F^{(0)}] + i\, {\rm Im}\, \mathcal{S}_{\pm}[F^{(0)}] + \sigma \epsilon^{-1} e^{-A/\epsilon } \mathcal{S}_\pm [F^{(1)}]\ . \label{eq8}
\end{align}
The second term corresponds to the imaginary ambiguity.  One can wisely choose $A$ and $\sigma$ in such a way that the third term cancels with the imaginary ambiguity and that the lateral Borel resummations become real-valued.  Let us write such $A$ and $\sigma$ as
\begin{align}
	A = \bar{A}\ ,\
	{\rm Im}\, \sigma = \pm \bar{\sigma}_I \ .
\end{align}
Note that ${\rm Re}\,\sigma$ cannot be fixed by the above cancellation and reality conditions; it is fixed by other conditions, i.e., initial/boundary conditions.  One can uniquely fix the values of $\bar{A}, \bar{\sigma}_I$ by examining the singularity structure of the Borel transform ${\mathcal B}[F^{(0)}]$.  For example, let us consider a Borel transform ${\mathcal B}[F^{(0)}]$ having a simple pole on the positive real axis $s=s_0>0$, whose imaginary ambiguity reads $ {\rm Im}\, \mathcal{S}_{\pm}[F^{(0)}] = \mp \pi  e^{-s_0/\epsilon}{\rm Res}\,\mathcal{B}[F^{(0)}](s_0)/\epsilon$.  One can cancel this imaginary ambiguity and make the Borel resummations (\ref{eq8}) real-valued if and only if one identifies $\bar{A}=s_0$ and $\bar{\sigma}_I = \pi {\rm Res}\, \mathcal{B}[F^{(0)}](s_0) / {\rm Re}\, \mathcal{S}_\pm [F^{(1)}]$.  If the third term $\mathcal{S}_\pm [F^{(1)}]$ is Borel summable, then the imaginary ambiguity is completely removed just by taking into account the one-instanton sector as above.  In general, however, the third term can be non-Borel summable, giving rise to another imaginary ambiguity of the order of ${\mathcal O}(e^{-2\bar{A}/\epsilon})$.  Such a higher order ambiguity can be canceled out precisely by including the $n=2$ instanton sector with the same $\bar{A}$ and $\bar{\sigma}_I$, but it can again induce another imaginary ambiguity of the order of ${\mathcal O}(e^{-3\bar{A}/\epsilon})$.  In general, this loop continues indefinitely, but one can cancel out all the imaginary ambiguities by successively incorporating all the multi-instanton sectors, with the same $\bar{A}$ and $\bar{\sigma}_I$ determined in the one-instanton sector.  This is because multi-instanton sectors in the trans-series are connected with one another, which is the main prediction of the resurgence theory \cite{AnicSchi15:NonpeAmbigandRealiResurTrans}.  In this way, one can obtain a unique real-valued function via the following {\it median resummation} $\mathcal{S}_{\rm med}$ defined by
\begin{align}
	&\mathcal{S}_{\rm med} [\tilde{F}] \nonumber\\
	&:= \mathcal{S}_\pm [\tilde{F}(\epsilon; \bar{A}, {\rm Re}\,\sigma \mp i \bar{\sigma}_I)]  \nonumber\\
	&= \frac{\mathcal{S}_-[\tilde{F}(\epsilon; \bar{A}, {\rm Re}\,\sigma+i \bar{\sigma}_I )] + {\mathcal S}_+[\tilde{F}(\epsilon; \bar{A}, {\rm Re}\,\sigma -i \bar{\sigma}_I )]}{2} \nonumber\\
	&= {\rm Re} \frac{\mathcal{S}_-[F^{(0)}] + {\mathcal S}_+[F^{(0)}] }{2} + {\mathcal O}( e^{-\bar{A}/\epsilon}) \ . \label{-eq9}
\end{align}
The ${\mathcal O}( e^{-\bar{A}/\epsilon})$-correction is absent for Borel transforms having only isolated poles on the real axis, for which case the median resummation $\mathcal{S}_{\rm med}$ exactly reduces to the average of the two lateral Borel resummations (or taking the Cauchy principle value). Note that ${\mathcal S}_{\rm med} = {\mathcal S}$ if there are no singularities on the real axis.

Having explained the basics of the Borel resummation technique, we introduce several terminologies of hydrodynamics in terms of the Borel resummation.  In hydrodynamics, an observable $F$ is a bulk quantity such as energy density and pressure, and the small parameter $\epsilon$ is identified with gradient (e.g., in kinetic description, $\epsilon = \mathrm{Kn}$ with ${\rm Kn}$ being the Knudsen number defined as the ratio of the typical mean-free path of microscopic processes to the typical macroscopic length-scale over which bulk quantities vary).  Hydrodynamics is  conventionally defined as a low-order truncation of the gradient expansion near equilibrium, where all the non-perturbative effects have been omitted \cite{Romatschke:2017ejr}. Namely, $F^{(0)}_0$ corresponds to ideal hydrodynamics, governed by the Euler equation, and $\sum_{k=0}^{\kappa} F^{(0)}_{k}$ ($\kappa \geq 1$) corresponds to the $\kappa$-th order viscous hydrodynamics (e.g., $\kappa=1$ is for Navier-Stokes hydrodynamics).  The (median) Borel resummation ${\mathcal S}_{\rm med}[\tilde{F}]$ contains all the $k$-th order gradients as well as non-perturbative effects that cannot be captured within the naive perturbative expansion without instantons.  The non-perturbative effects are included in nonzero $n$-instanton sectors of the Borel resummation, i.e., the second term of ${\mathcal S}_{\rm med}[\tilde{F}] = {\mathcal S}_{\rm med}[F^{(0)}] + \sum_{n=1}^{\infty} \sigma^n \epsilon^{-n} e^{-nA/\epsilon} {\mathcal S}_{\rm med}[F^{(n)}]$, which we call the $n$-th order {\it non-hydrodynamic modes}.  The non-hydrodynamic modes decay with typical time/length-scale set by the instanton action $A$ (times the number of instantons $n$).  In the long-wavelength/low-frequency limit, therefore, only the first term ${\mathcal S}_{\rm med}[F^{(0)}]$ survives and asymptotes the low-order hydrodynamics such as $F^{(0)}_0$ and $F^{(0)}_1$, and hence we call ${\mathcal S}_{\rm med}[F^{(0)}]$ the {\it hydrodynamic attractor}.

\section{Hydrodynamic analysis} \label{sec:3}

\subsection{MIS theory under a Hubble expansion} \label{sec:3a}

We consider a Hubble expansion, i.e., a spatially homogeneous and isotropic expansion in three dimensions.  The metric reads
\begin{align}
	ds^2=dt^2-a^2(t)\left(dx^2+dy^2+dz^2\right) \ , \label{eq12}
\end{align}
where $a(t)$ is a dimensionless scale factor, controlling the speed of the expansion, and we assumed for simplicity that the curvature parameter is vanishing.  For simplicity, we assume that the scale factor $a$ has the following power-type dependence on $t$, rather than being determined by the Einstein equation,
\begin{align}
	a(t) = (t/t_{\rm in})^\alpha\ ,  \label{eq13}
\end{align}
with some initialization time $t_{\rm in}$.  That is to say, our expansion is manually controlled, and energy is continuously injected into or extracted from the system.  The choice of the form (\ref{eq13}) for $a$, albeit manually, has a close relation to cosmology.  In fact, the Friedmann equation for a flat universe of a single ingredient always yields a scale factor $a$ in the power-law type.  In the following, we will take the value of the exponent $\alpha$ in Eq.~(\ref{eq13}) to be $2/3$, which is inspired by the scale factor in a matter-dominated universe \cite{Carroll}.  Moreover, the manually controlled expansion may be realized in cold atomic systems where the expansion rate can be controlled by tuning the trapping potential.  In this case, the non-hydrodynamic modes may appear during the dynamical expansion of the system~\cite{schafer2014,brewer2015}.

We consider a viscous fluid whose dynamics is governed by the MIS theory \cite{Muller:1967zza,ISRAEL1979341}.  The MIS theory consists of the energy-momentum conservation equation and a relaxation type equation for bulk stress.  Unlike the NS theory, the equations in the MIS theory restore the relativistic causality and stability and fluid variables relax to the NS results in the late-time limit.  In the local rest frame of a fluid under the Hubble expansion (\ref{eq12}), the MIS equations read
\begin{subequations}
\label{eq:EOM}
\begin{align}
	&\dot{E} + 3\frac{\alpha}{t}(P+E+\Pi) = 0\ ,  \label{eq:EOM1} \\
	&\tau_\Pi \dot{\Pi} + \Pi = -3\zeta\frac{\alpha}{t}\ , \label{eq:EOM2}
\end{align}
\end{subequations}
where the dot denotes the time derivative and we used $\dot{a}/a = \alpha/t$ for the manual expansion (\ref{eq13}).  The shear viscous corrections are vanishing because of the spatial isotropy of the system.  $E, P, \Pi, \zeta$, and $\tau_\Pi$ are energy density, pressure, bulk stress, bulk viscosity, and relaxation time, respectively.  Equations~(\ref{eq:EOM}) form a closed set when an equation of state $P=P(E)$ is specified.  In this paper, we use
\begin{align}
	P=c_s^2E\ , \label{eq15}
\end{align}
with $c_s$ being the speed of sound.  We treat the transport coefficients $\zeta,\tau_\Pi$ and the speed of sound $c_s$ as constants in the following calculations.

\subsection{Hydrodynamic attractor} \label{sec:3b}

To get a hydrodynamic attractor solution in the viscous system under the manual Hubble expansion, we first demonstrate how factorial divergences appear in the naive perturbative expansion without instanton corrections of the bulk stress $\Pi^{(0)}$ and the energy density $E^{(0)}$.  The relaxation time $\tau_\Pi$ is the typical time for the damping of the bulk stress towards the NS result.  The time derivative term in Eq.~(\ref{eq:EOM2}) describes the deviation of the bulk stress from the corresponding NS result.  Therefore, the bulk stress keeps away from the NS result for a longer time with larger $\tau_\Pi$, so we treat the ratio $\tau_\Pi/t$ as a perturbative parameter (will be simply called ``gradient'').  In this way, we will expand hydrodynamic variables around the NS theory~\footnote{One can adopt an alternative assignment to expand around the ideal fluid hydrodynamics by treating both $\tau_\Pi$ and $\zeta$ as perturbation, but the results only differ by one order of $\epsilon$.}.  To make the perturbative expansion more tractable, we introduce a dimensionless book-keeping parameter $\epsilon$, which will be taken to be unity after the whole calculation is completed, to the original equation (\ref{eq:EOM2}) as
\begin{equation}
	\epsilon \tau_\Pi \dot{\Pi} + \Pi = -3\zeta\frac{\alpha}{t}\ . \label{eq16}
\end{equation}

We expand the bulk stress $\Pi^{(0)}$ and the energy density $E^{(0)}$ in terms of the book-keeping parameter $\epsilon$ as
\begin{subequations}
\label{eq17}
\begin{align}
	E^{(0)}(t;\epsilon)
		:= \sum_{k=0}^\infty E^{(0)}_k(t) \epsilon^k\ , \\
	\Pi^{(0)}(t;\epsilon)
		:=\sum_{k=0}^\infty \Pi^{(0)}_k(t) \epsilon^k \ .
\end{align}
\end{subequations}
Substituting Eqs.~(\ref{eq17}) into Eqs.~(\ref{eq:EOM}), we find

\begin{subequations}
\label{eq:pertsol}
\begin{align}
		\Pi^{(0)}_k
		&= -3\zeta \frac{\alpha}{t} \Gamma(k+1) \left( \frac{\tau_\Pi}{t} \right)^{k}\ , \label{eq:pertsolPi} \\
	E^{(0)}_k	
		&=-3\zeta \frac{\alpha}{t} \frac{3\alpha\Gamma(k+1)}{k+1-Q} \left( \frac{\tau_\Pi}{t} \right)^{k} \ , \label{eq:pertsolE}
\end{align}
\end{subequations}
where $k \geq 0$ and
\begin{align}\label{eq20}
	Q := 3\alpha(c_s^2+1)\ .
\end{align}
We assume $1-Q<0$ [which in turn sets a lower limit on the expansion rate $\alpha$ as $\alpha>(3(c_s^2+1))^{-1}$], so that the energy density $E^{(0)}$ is positive on the first order \footnote{When Eq.~(\ref{eq:EOM1}) is considered as an inhomogeneous differential equation for $E$, $E={\rm const.} \times t^{-Q}$ is its complementary solution.  This part is irrelevant to the perturbation program, and as we will see in Sec. \ref{sec:3b}, it will show up in the first non-hydrodynamic mode after resummation.}.  It is evident that the coefficients in the perturbative solution (\ref{eq:pertsol}) diverge factorially.

Let us apply the Borel resummation to the perturbative solution (\ref{eq:pertsol}) to obtain a hydrodynamic attractor.  We first look into the bulk stress $\Pi^{(0)}$, whose Borel transform reads
\begin{align}\label{eq:Borelpi}
	\mathcal{B}[\Pi^{(0)}](s)
		= -3\zeta \frac{\alpha}{t} \frac{1}{1-s \frac{\tau_\Pi}{t}}\ .
\end{align}
The Borel transform of $\Pi^{(0)}$ has a simple pole at $s=t/\tau_\Pi$ on the real axis, and it gives rise to a non-perturbative imaginary ambiguity through the lateral Borel resummation (\ref{eq5}) as
\begin{align}
	{({\mathcal S}_+ - {\mathcal S}_-)[\Pi^{(0)}]
		= - 6\pi i \zeta \frac{\alpha}{t} \frac{t}{\tau_\Pi} e^{-\frac{t}{\tau_\Pi}}}\ , \label{eq22}
\end{align}
where and hereafter we take $\epsilon = 1$ and suppress $\epsilon$ unless it is needed.  Note that there are no higher-order imaginary ambiguities [i.e., no higher-order instanton effects $\mathcal{O}(e^{-nt/\tau_\Pi}), n \geq 2$], implying the absence of $n$th-order non-hydrodynamic modes with $n\geq 2$.  This point will be discussed in more detail in Sec.~\ref{sec:3d}.  One can remove the imaginary ambiguity by taking the median resummation, as we have explained in Sec.~\ref{sec:2}, and obtain a hydrodynamic attractor solution for the bulk stress $\Pi$ as
\begin{align}\label{eq:resumPi}
	{\mathcal{S}_{\rm med}[\Pi^{(0)}]
		= -3\zeta \frac{\alpha}{t} \frac{t}{\tau_\Pi} e^{-\frac{t}{\tau_\Pi}}\mathrm{Ei}\left( \frac{t}{\tau_\Pi} \right)} \ ,
\end{align}
where $\mathrm{Ei}$ denotes the exponential integral function. The attractor solution (\ref{eq:resumPi}) contains all-order gradients $(\tau_\Pi/t)^k$ and behaves in the limit of small gradient $\tau_\Pi \to 0$ (or late-times $t \to \infty$) as
\begin{align}\label{eq24}
	{\mathcal{S}_{\rm med}[\Pi^{(0)}]
		= -3\zeta \frac{\alpha}{t} \left[ 1 + \frac{\tau_\Pi}{t} + 2 \left( \frac{\tau_\Pi}{t} \right)^2 + {\mathcal O}\left( \left( \frac{\tau_\Pi}{t} \right)^3 \right) \right]}\ ,
\end{align}
which is in full agreement with the perturbative solution (\ref{eq:pertsolPi}), i.e., the hydrodynamic attractor (\ref{eq:resumPi}) reduces to the standard low-order hydrodynamics at late times.

Next, we turn to resum the energy density $E^{(0)}$.  The Borel transform is given in terms of a hypergeometric function ${}_2F_1$ as
\begin{align}\label{eq25}
	{{\mathcal B}[E^{(0)}](s)
		= -3\zeta \frac{\alpha}{t} \frac{3\alpha}{1-Q} {{}_2F_1}\left({1-Q,1\atop 2-Q};\frac{s\tau_\Pi}{t}\right)}\ ,
\end{align}
which has a branch cut extending from $s=t/\tau_\Pi$ to the positive infinity.  It leads to a non-perturbative imaginary ambiguity,
\begin{align}
	{({\mathcal S}_+ - {\mathcal S}_-)[E^{(0)}]}
		&{= -18 \alpha \pi i \zeta \frac{\alpha}{t}  \left(\frac{t}{\tau_\Pi}\right)^{1-Q} \Gamma \left( Q, \frac{t}{\tau_\Pi} \right)} \nonumber\\
		&{= -18 \alpha \pi i \zeta \frac{\alpha}{t}   e^{-\frac{t}{\tau_\Pi}} \left[ 1 + {\mathcal O}\left(\frac{\tau_\Pi}{t} \right) \right]} \ , \label{eq26}
\end{align}
where $\Gamma(x,t)$ is the incomplete Gamma function.  Note that, similar to the bulk stress (\ref{eq22}), we have no instanton contributions of the order ${\mathcal O}(e^{-2t/\tau_\Pi})$ to the energy density $E$.  With the median resummation, one can kill the imaginary ambiguity and obtain a hydrodynamic attractor solution for the energy density $E$,
\begin{align}\label{eq:resumE}
	{{\mathcal S}_{\rm med }[E^{(0)}]}
		&{= -3\zeta \frac{\alpha}{t} \frac{3 \pi \alpha}{\sin (\pi Q)} \frac{t}{\tau_\Pi} \Biggl[ G^{2,1}_{2,3} \left( \left.\frac{t}{\tau_\Pi} \right| \begin{array}{l} 0;1- Q \\ 0,0;-Q\end{array} \right)} \nonumber\\
		&{\quad - \cos(\pi Q) \left(\frac{t}{\tau_\Pi}\right)^{-Q} \Gamma \left( Q, \frac{t}{\tau_\Pi} \right) \Biggl]} \ .
\end{align}
where $G$ is Meijer's G-function.  One can confirm that the attractor solution (\ref{eq:resumE}) correctly asymptotes the standard low-order hydrodynamics, or the perturbative solution (\ref{eq:pertsolE}), as
\begin{align}
	{{\mathcal S}_{\rm med}[E^{(0)}]}
		&{= -3\zeta \frac{\alpha}{t} \left[ \frac{3\alpha}{1-Q} + \frac{3\alpha}{2-Q} \frac{\tau_\Pi}{t} \right.} \nonumber\\
		&{\quad \left. +  \frac{6\alpha}{3-Q} \left( \frac{\tau_\Pi}{t} \right)^2  + {\mathcal O}\left(  \left( \frac{\tau_\Pi}{t} \right)^3 \right) \right]} \ , \label{eq28}
\end{align}

\begin{figure*}
\includegraphics[width=0.45\textwidth]{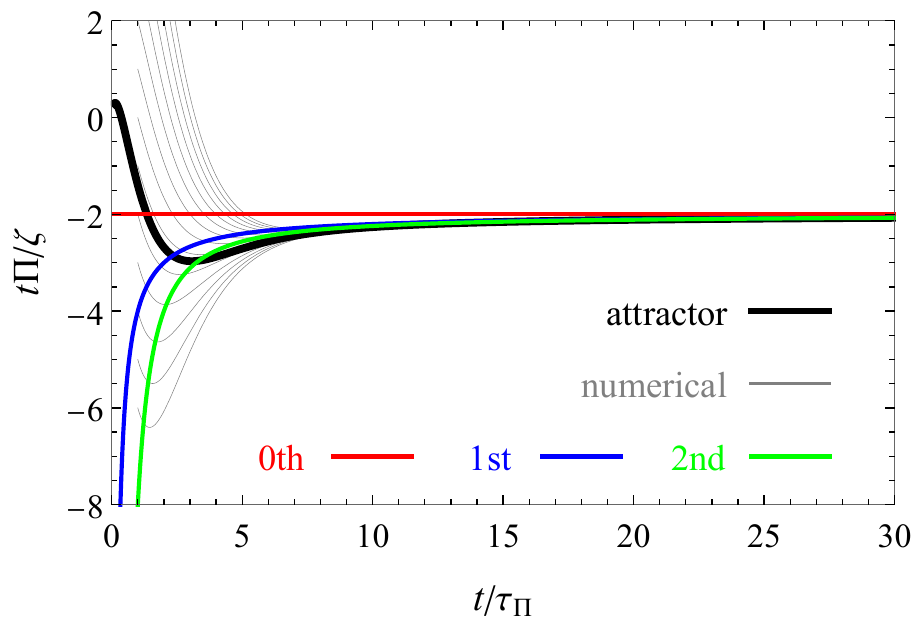}\hspace{8mm}
\includegraphics[width=0.45\textwidth]{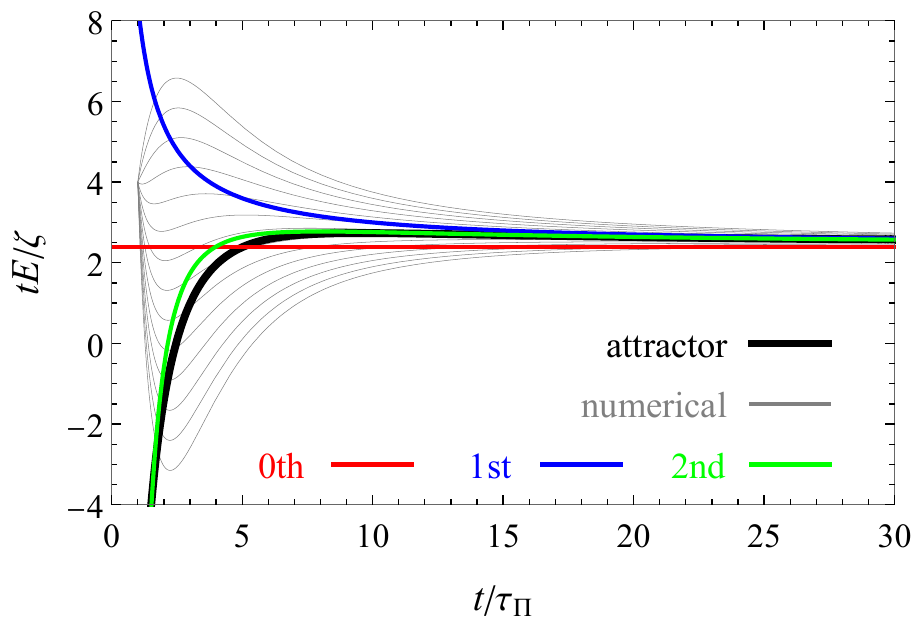}
\caption{The hydrodynamic attractor solutions (black lines) for the bulk stress (\ref{eq:resumPi}) (left panel) and the energy density (\ref{eq:resumE}) (right panel) as functions of time $t/\tau_\Pi$.  As a comparison, we also plot numerical solutions to the MIS theory (\ref{eq:EOM}) with various initial conditions (gray lines) and the low-order hydrodynamics (\ref{eq:pertsol}) of the zero-th (red line), first (blue line), and second order (green line).  The parameters are set as $\alpha=2/3$ and $c_s^2=1/3$.  The initial conditions for the numerical solutions are set at $t_{\rm in}/\tau_\Pi=1$ as $t_{\rm in}\Pi(t_{\rm in})/\zeta=-7+m$ and $t_{\rm in}E(t_{\rm in})/\zeta=4$ with $m=1,2,\cdots,14$.}
\label{fig:att1}
\end{figure*}

Figure~\ref{fig:att1} shows the hydrodynamic attractor solutions for the bulk stress (\ref{eq:resumPi}) and the energy density (\ref{eq:resumE}).  As a comparison, we also plotted the perturbative solutions (\ref{eq:pertsol}), i.e., the low-order hydrodynamics, and numerical solutions of the MIS theory (\ref{eq:EOM}) for a fixed value of $E_{\rm in}$ but various initial bulk stress $\Pi_{\rm in}$.  As is shown in Fig.~\ref{fig:att1}, the numerical solutions with different initial conditions universally converge to the hydrodynamic attractor solutions well before the low-order hydrodynamics' work at late times, where the gradient becomes small.  Such a convergent behavior appears because the information of the initial conditions is basically taken into account by non-hydrodynamic modes, which we discuss in the subsequent subsection.  The non-hydrodynamic modes are short-lived and decay with a typical time-scale $t \sim \tau_\Pi$, after which the dynamical evolution of the system is universally described by the hydrodynamic attractor solutions.

We comment that both attractor solutions for $\Pi$ and $E$ are globally attractive.  To see this, let us make variations $\delta \Pi, \delta E$ around some known solution to Eqs.~(\ref{eq:EOM}), or the hydrodynamic attractor solutions (\ref{eq:resumPi}) and (\ref{eq:resumE}), as
\begin{subequations}\label{eq27}
\begin{align}
	&\frac{d}{dt}{\delta E} + \frac{1}{t}(Q \delta E+3 \alpha \delta \Pi) = 0\ ,  \label{eq:27a} \\
	&\tau_\Pi \frac{d}{dt} \delta \Pi + \delta \Pi = 0\ , \label{eq:27b}
\end{align}
\end{subequations}
with the initial conditions
\begin{subequations}\label{eq27i}
\begin{align}
	&\delta \Pi(t_{\rm in})=:\delta \Pi _{\rm in} \ ,  \label{eq:27ia} \\
	&\delta E(t_{\rm in}) =: \delta E_{\rm in}\ . \label{eq:27ib}
\end{align}
\end{subequations}
The solutions for  $\delta \Pi, \delta E$ are,
\begin{subequations}\label{eq23}
\begin{align}
	&\delta \Pi(t)=\delta \Pi _{\rm in} e^{-\frac{t-t_{\rm in}}{\tau_\Pi}}\ ,  \label{eq:23a} \\
	&\delta E(t) = -3\alpha\left(\frac{\tau_\Pi}{t}\right)^Q e^{\frac{t_{\rm in}}{\tau_\Pi}}(\Gamma(Q,t_{\rm in})-\Gamma(Q,t))\delta \Pi_{\rm in}\nonumber\\
&\qquad\quad+\left(\frac{t_{\rm in}}{t}\right)^Q \delta E_{\rm in}\ . \label{eq:23b}
\end{align}
\end{subequations}
From Eq.~(\ref{eq23}), one may observe that $E$ approaches the attractor solution slowly at a rate of the power law, while the deviation in $\Pi$ decays exponentially and thus it is rapid.  This observation is consistent with Fig.~\ref{fig:att1}, in which the numerical solutions for $\Pi$ converge faster than the ones for $E$ with the same set of initial values.  Note that $\delta E$ decays faster with increasing $Q$, i.e., larger expansion rate $\alpha$, while the decay speed is controlled only by $\tau_\Pi$ for the bulk stress $\delta \Pi$.  It is worth stressing that since there are no restrictions on the values of $\delta \Pi_{\rm in}$ and $\delta E_{\rm in}$, the attractor solutions we obtained are globally attractive.

\subsection{Resummed viscosity coefficient} \label{sec:3c}

The hydrodynamic attractor solution of the bulk stress (\ref{eq:resumPi}) may be reshaped into the form of a Navier-Stokes constitutive relation [i.e., $\tau_\Pi \to 0$ limit of Eq.~(\ref{eq:EOM2})] with an effective viscosity coefficient $\zeta_B$:
\begin{align}\label{eq29}
	{\mathcal S}_{\rm med}[\Pi^{(0)}] =: - 3\zeta_B \frac{\alpha}{t}\ .
\end{align}
From the explicit form of the hydrodynamic attractor solution (\ref{eq:resumPi}), one finds that the effective viscosity $\zeta_B$ is given as a function of the gradient strength $\tau_\Pi/t$ as
\begin{align}
	\frac{\zeta_B}{\zeta}
		= \frac{t}{\tau_\Pi} e^{-\frac{t}{\tau_\Pi}}\mathrm{Ei}\left(\frac{t}{\tau_\Pi} \right)\ . \label{eq30}
\end{align}
Therefore, $\zeta_B$ includes higher-order gradient effects and asymptotes the original bare viscosity $\zeta$ at late times, where the gradient becomes small, as
\begin{align}
	\frac{\zeta_B}{\zeta} = 1 + \frac{\tau_\Pi}{t} + {\mathcal O}\left( \left( \frac{\tau_\Pi}{t} \right)^2 \right)\ .  \label{eq31}
\end{align}

\begin{figure}
\includegraphics[width=0.45\textwidth]{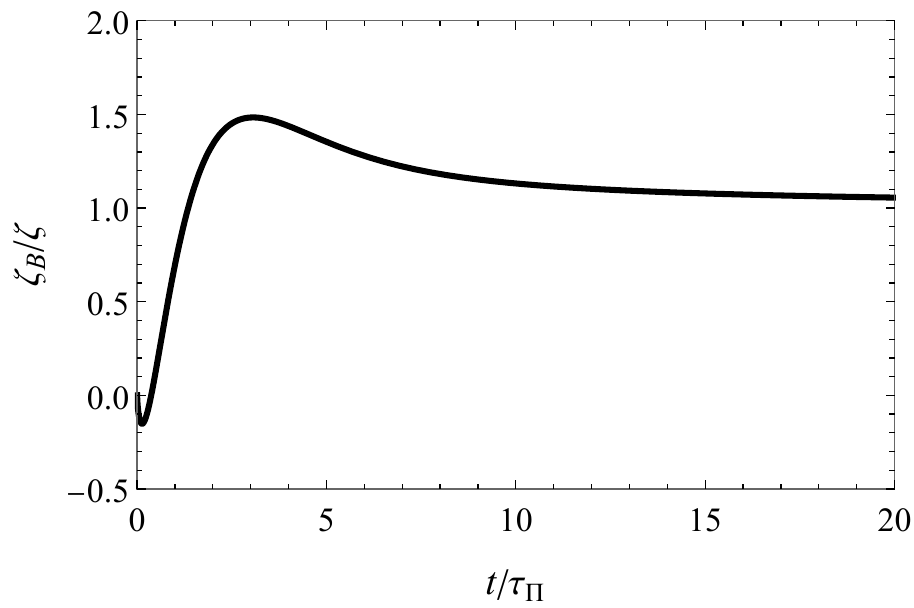}
\caption{The effective viscosity coefficient $\zeta_B$ (\ref{eq30}) versus time.}
\label{fig:viseff2}
\end{figure}

Figure~\ref{fig:viseff2} shows the effective viscosity $\zeta_B$ (\ref{eq30}) plotted against time, or the inverse of the gradient strength $t/\tau_\Pi$.  In the late-time limit or small gradient $t/\tau_\Pi \to \infty$, high-order gradient corrections to the bulk viscosity are strongly suppressed and only the Navier-Stokes viscosity coefficient revives, as anticipated in Eq.~(\ref{eq31}).  For intermediate values of $t/\tau_\Pi$, high-order gradient corrections manifest themselves as the enhancement of $\zeta_B$ with a peak reached at $t/\tau_\Pi \sim 3$.  The extremum of the effective  viscosity in this parameter regime appears also in resummed Baier-Romatschke-Son-Starinets-Stephanov(rBRSSS) hydrodynamics~\cite{Baier_2008} under the Bjorken flow~\cite{romatschke_2}.  In the early stage of the expansion when $t/\tau_\Pi\lesssim 1$ is small, one can observe that the effective bulk viscosity is much lowered than the bare value $\zeta$.  This tendency is similar to the effective shear viscosity of rBRSSS hydrodynamics and the kinetic theory under the Bjorken flow~\cite{romatschke_2, Behtash:2021coeffRG,Behtash:2018moe} and indicates that dissipative effects may be suppressed under non-equilibrium situations~\cite{Lublinsky:2007mm,Bu:2014sia}.  While the effective bulk viscosity stays positive in a large regime of gradients, it can be negative at the very early times $t/\tau_\Pi < 0.373$, where the gradient is extremely large and hydrodynamic picture may not be applicable.  Negative effective bulk viscosity indicates a positive modification on pressure and decrease in entropy, and so energy needs to be injected into the system for keeping the manual expansion process.

\subsection{Non-hydrodynamic mode}\label{sec:3d}

We discuss non-hydrodynamic contributions to $E$ and $\Pi$ using the trans-series ansatz.  We show that only the first non-hydrodynamic mode can contribute in our manual Hubble-expanding system.

We begin with assuming the trans-series ansatz:
\begin{subequations}
\label{eq32}
\begin{align}
	\tilde{\Pi}(t;\epsilon; \sigma)	
		&:= \sum_{n=0}^{\infty} \sigma^n {\epsilon^{-n}} e^{-n \frac{t}{\epsilon \tau_\Pi}} \Pi^{(n)}(t;\epsilon)\ , \\
	\tilde{E}(t;\epsilon;\sigma)	
		&:= \sum_{n=0}^{\infty} \sigma^n {\epsilon^{-n}} e^{-n \frac{t}{\epsilon  \tau_\Pi}} E^{(n)}(t;\epsilon)\ ,
\end{align}
\end{subequations}
where we have recovered the book-keeping parameter $\epsilon$ for clarity and identified the instanton action with $t/\epsilon \tau_\Pi$ from the magnitude of the imaginary ambiguities (\ref{eq22}) and (\ref{eq26}).  The perturbative expansions around $n$-instanton background $\Pi^{(n)}$ and $E^{(n)}$ are defined as
\begin{subequations}
\begin{align}\label{eq33}
	E^{(n)}(t;\epsilon) := \sum_{k=0}^\infty E^{(n)}_k(t) \epsilon^k\ , \\
	\Pi^{(n)}(t;\epsilon) := \sum_{k=0}^\infty \Pi^{(n)}_k(t) \epsilon^k \ .
\end{align}
\end{subequations}
Substituting the trans-series ansatz (\ref{eq32}) into the equations of motion (\ref{eq:EOM}), one obtains recursive relations for the coefficients $\Pi^{(n)}_k$ and $E^{(n)}_k$ as
\begin{subequations}
\label{eq:recPi}
\begin{align}
	0 &= (1-n)\Pi^{(n)}_{0} \ ,\\
	0 &= (1-n)\Pi^{(n)}_{k+1} + \tau_\Pi \dot{\Pi}^{(n)}_{k} \ ,  \label{eq60b}
\end{align}
\end{subequations}
and
\begin{subequations}
\label{eq:recE}
\begin{align}
	0 &= n E_0^{(n)} \ ,\\
	0 &= n E_{k+1}^{(n)} - \tau_\Pi \left[ \dot{E}^{(n)}_k + \frac{Q}{t} E_k^{(n)} + 3\frac{\alpha}{t} \Pi_k^{(n)}  \right] \ , \label{eq61b}
\end{align}
\end{subequations}
where $n\geq 1$ and $k\geq 0$.

From Eq.~(\ref{eq:recPi}), one may observe that all the $\Pi^{(n)}_k$'s have to vanish except for $\Pi^{(1)}_k$'s, i.e., only the first non-hydrodynamic mode can contribute to $\Pi$.  Furthermore, one can show that
\begin{align}
	0 = \Pi^{(1)}_k\ {\rm for}\ k\geq 1\ , \label{eq36}
\end{align}
and only $\Pi^{(1)}_0$ can be non-vanishing.  To see this, we remind that the book-keeping parameter $\epsilon$ is introduced via the replacement $\tau_\Pi \to \epsilon \tau_\Pi$, so that $\Pi^{(1)}_k$ must contain $k$ folds of $\tau_\Pi$'s.  Also, putting aside the bulk viscous constant $\zeta$, which can enter in $\Pi^{(1)}$ just as an overall factor, we only have two dimensionful quantities $t$ and $\tau_\Pi$, out of which we can have only one dimensionless combination $\tau_\Pi/t$.  Those observations imply $\Pi^{(1)}_k = C_k (\tau_\Pi/t)^{k}$, with $C_k$ being some time-independent constant.  On the other hand, the recursive relation (\ref{eq60b}) tells us that $0 = \dot{\Pi}^{(1)}_k = -k \tau_\Pi C_k (\tau_\Pi/t)^{-k-1}$.  Therefore, we have $C_k=0$ for $k\geq 1$, which leads to Eq.~(\ref{eq36}).  The remaining non-vanishing constant $\Pi^{(1)}_0$ is determined by initial/boundary conditions for the bulk stress $\Pi$.

Similarly, we find from Eq.~(\ref{eq:recE}) that only $E_k^{(1)}$ can be non-vanishing and hence only the first non-hydrodynamic mode can contribute to $E$.  Unlike $\Pi^{(1)}_k$, $E^{(1)}_k$ has non-trivial $k$-dependence and gives a factorial divergence.  Indeed, one can explicitly solve the recursive relation (\ref{eq61b}) and find
\begin{align}
	E^{(1)}
		\simeq C \sum_{k=0} \left(- \epsilon\frac{\tau_\Pi}{t} \right)^k \frac{\Gamma(k-Q)}{\Gamma(-Q)} (1 - \delta_{k,0}) \ , \label{eq38}
\end{align}
with $C$ being an integration constant that is fixed by initial/boundary  conditions, as in the case of $\Pi^{(1)}_0$.  To make the formal power series $E^{(1)}$ well-defined, we apply the Borel resummation.  The Borel transform reads
\begin{align}\label{eq40}
	{\mathcal B}[E^{(1)}] = C \left[ \left( 1 + s \frac{\tau_\Pi}{t}  \right)^Q - 1 \right].
\end{align}
Importantly, the Borel transform ${\mathcal B}[E^{(1)}]$ does not possess any singularities on the positive real axis.  Hence, it is Borel summable and is free from the imaginary ambiguity.  The absence of the imaginary ambiguity is essentially related to the absence of the higher-order non-hydrodynamic modes; if it exists, higher-order non-hydrodynamic modes should exist and vice versa.  With the Borel resummation, one obtains a well-defined function describing the first non-hydrodynamic contribution to the energy density ${\mathcal S}[E^{(1)}]$ as
\begin{align}\label{eq41}
	{\mathcal S}[E^{(1)}]
		=  C \left[ \left( \frac{\tau_\Pi}{t} \right)^{Q} e^{\frac{t}{\tau_\Pi}} \Gamma\left( 1+Q, \frac{t}{\tau_\Pi}\right) - 1 \right] \ ,
\end{align}
where $\epsilon \to 1$ is understood.

\section{Kinetic analysis}\label{sec:4}

We explore how hydrodynamic attractors appear in kinetic theory and compare them with the MIS result.

Our starting point is a Boltzmann equation with relaxation-time approximation in the Hubble spacetime:
\begin{equation}\label{eq:Boltzmann}
	\left[ \frac{\partial}{\partial t} -\frac{2\dot{a}}{a}p\frac{\partial}{\partial p} \right] f{(t,p)}
	= \frac{{f^{(0)}_0}{(t,p)}-f{(t,p)}}{\tau_R}\ ,
\end{equation}
where $\tau_R$ is relaxation time and we assume a massless limit for simplicity so that single-particle energy $p^0$ equals the local momentum $p^0 = pa$ (with $p:=|\vec{p}|$)~\footnote{The on-shell condition in the Hubble metric is $p_0^2-(pa)^2=m^2$, so $pa$ is interpreted as the local momentum.}.  $f$ and $f^{(0)}_0$ are, respectively, a single-particle distribution function and its ``equilibrium" form to which $f$ relaxes after the time evolution~\footnote{At $f=f^{(0)}_0$, there are no microscopic collisions [i.e., the right-hand side of the Boltzmann equation (\ref{eq:Boltzmann}) vanishes] and $f^{(0)}_0$ is thus called a local-equilibrium distribution.  On the other hand, such $f^{(0)}_0$ does not necessarily vanish the left-hand side of Eq.~(\ref{eq:Boltzmann}), meaning that the system is not globally equilibrated due to expansion and that $f=f^{(0)}_0$ cannot be a solution except for the limit $\tau_R \to 0$.  Nevertheless, for $\alpha<3/4$ in Eq.~(\ref{eq13}) and $T(t)$ given in Eq.~(\ref{eq54}), the global equilibrium is reached at $t\rightarrow\infty$.  This is actually the situation that we consider in the numerical calculations.}.  We assume that $f^{(0)}_0$ takes a Maxwellian form
\begin{align}
	f^{(0)}_0(t,p) := \exp(-pa(t)/T(t)) \ ,
\end{align}
where $T$ can be interpreted as a parameter that characterizes temperature of the system.  At this stage, $T$ can be an arbitrary function of $t$ and its explicit form is fixed by imposing some matching condition and/or physical requirements, as we will do later.

One may obtain a general solution to  Eq.~(\ref{eq:Boltzmann}) using the method of characteristics.  The solution reads
\begin{align}\label{eq43}
	f(t,p)	&= \frac{1}{\tau_R} \int_{t_{\rm in}}^t  dt' e^{-\frac{t-t'}{\tau_R}} e^{ -\frac{pa^2(t)}{a(t')T(t')} }\nonumber\\
				&\quad + f_{\rm in}(pa^2(t)) e^{-\frac{t-t_{\rm in}}{\tau_R} } \ ,
\end{align}
with $f_{\rm in}(p) := f(t_{\rm in},p)$ being the initial distribution at $t=t_{\rm in}$.  One may interpret the general solution of $f$~(\ref{eq43}) as a sum of ``a hydrodynamic generator'' (first term) and a source for non-hydrodynamic modes (second term), since only the second term can contain information of initial conditions and hydrodynamic attractors appear when the initial information is lost; similar discussions can be found in Ref.~\cite{McNelis:2020:Hgrkt} for the case of the Bjorken flow.  Using Eq.~(\ref{eq43}), we can evaluate the energy density as the first-order moment of the distribution function,
\begin{align}\label{eq45}
	E(t)	&:= \int \frac{d^3\vec{p}}{(2\pi)^3} \sqrt{-g(t)} pa(t) f(t,p) \nonumber\\
 			&= \frac{3}{\pi^2} \frac{1}{a^4(t)} \int_{t_{\rm in}}^t \frac{dt'}{\tau_R} e^{-\frac{t-t'}{\tau_R}} a^4(t') T^4(t')\nonumber\\ &\quad + \frac{1}{a^4(t)} e^{-\frac{t-t_{\rm in}}{\tau_R}} E_{\rm in} \ ,
\end{align}
where $\sqrt{-g} = a^3$ is the determinant of the Hubble metric and $E_{\rm in}$ is the initial energy density at time $t=t_{\rm in}$, i.e.,
\begin{align}
	E_{\rm in}
		&:= \left. \int \frac{d^3\vec{p}}{(2\pi)^3} \sqrt{-g} pa(t) f(t,p) \right|_{t=t_{\rm in}} \nonumber\\
		&= \frac{1}{2\pi^2} \int_0^\infty dp\,p^3 f_{\rm in}(p) \ .
\end{align}
The energy density $E$ is time-dependent and is a decreasing function of $t$ because the spatial volume is increasing. At late times $t \to \infty$, the second term in Eq.~(\ref{eq45}) is vanishing and the integral in the first term may be dominated by contributions around $t'\sim t$ because of the exponential factor $e^{-(t-t')/\tau_R}$.  Thus, at the leading order in $\tau_R$, we get
\begin{align}
	E
		\xrightarrow[]{t \to \infty}\;& \frac{3}{\pi^2} \frac{1}{a^4(t)} \int_{t'\sim t} \frac{dt'}{\tau_R} e^{-\frac{t-t'}{\tau_R}} a^4(t) T^4(t) \nonumber\\
		\sim\;& \frac{3}{\pi^2} T^4(t)\ , \label{eq46}
\end{align}
which agrees with the contribution from $f^{(0)}_0$,
\begin{align}	
	E^{(0)}_0(t)
		&:= \int \frac{d^3\vec{p}}{(2\pi)^3} \sqrt{-g(t)} pa(t) f^{(0)}_0(t,p) \nonumber\\
		&= \frac{3}{\pi^2} T^4(t)\ ,
\end{align}
implying that the distribution function $f$ approaches $f^{(0)}_0$ at late times and that the information of the initial condition $f_{\rm in}$ is lost.  Note that the energy density deviates from the $T^4$-dependence (Stephan-Boltzmann law) except for late times due to the deviation $f-f^{(0)}_0 \neq 0$ and the resulting collision effects, i.e., the right-hand side of the Boltzmann equation (\ref{eq:Boltzmann}).

We turn to fix the functional form of $T$ by requiring a matching condition.  We first assume an ansatz that $T$ has the following power-type dependence parametrized by $T_{\rm in}$ and $q$,
\begin{align}
	T(t) =: \frac{ T_{\rm in} }{a^q(t)} \ , \label{eq54}
\end{align}
as the system cools down according to the power-type expansion (\ref{eq13}).  We then fix the parameters $T_{\rm in}$ and $q$ by matching the late-time behavior of the energy density in our kinetic description (\ref{eq46}) with that in NS hydrodynamics [$E^{(0)}_0$ in Eq.~(\ref{eq:pertsolE})].  We have
\begin{align}\label{kin:temp2}
	T_{\rm in} = \left| \frac{3\pi^2\alpha^2\zeta
}{(Q-1)t_{\rm in}} \right|^{1/4},\
	q = \frac{1}{4\alpha}\ .
\end{align}
Note that $Q-1>0$ [see the discussion below Eq.~(\ref{eq20})].

The right-hand side of the Boltzmann equation (\ref{eq:Boltzmann}) is responsible for microscopic collisions between particles and its strength is controlled by the relaxation time $\tau_R$.  We, therefore, regard $\tau_R$ (more precisely, gradient defined as the ratio of $\tau_R$ to $t$, which is the typical length scale for macroscopic quantities) as a small parameter and expand the single particle distribution $f$ perturbatively with regard to $\tau_R$ around its ``equilibrium" value $f^{(0)}_0$.  This expansion is known as the Chapman-Enskog (CE) expansion \cite{chapman53,CercignaniKremer} and is expressed as
\begin{align}\label{eq47}
	f &\simeq \sum_{k=0}^\infty \left[ -\epsilon\tau_R\left( \frac{\partial}{\partial t} -2\frac{\dot{a}}{a}p \frac{\partial}{\partial p}\right)\right]^{k} f^{(0)}_0 \nonumber\\
	  &=: \sum_{k=0}^\infty f^{(0)}_k \epsilon^k \nonumber\\
	  &=: f^{(0)}\ ,
\end{align}
where we have introduced a dimensionless variable $\epsilon$ as a book-keeping parameter and replaced $\tau_R \to \epsilon \tau_R$, as we did in Sec.~\ref{sec:3a}.
The CE expansion is, however, not convergent in general \cite{grad1963}.  It has been found to be factorially divergent in many situations \cite{CercignaniKremer,Denicol:2016bjh,Chen_2017}, so is in our case as we see below.

We show that the CE expansion leads to a factorially divergent series for the energy density in our Hubble-expanding system and how the Borel resummation can cure the divergence.  To this end, we explicitly evaluate the perturbative series for the distribution function  (\ref{eq47}) under the power-type assumptions for the scale factor $a$ given in Eq.~(\ref{eq13}) and temperature $T$ given in Eq.~(\ref{eq54}).  By changing the variables,
\begin{align}
	u := \frac{pa^2(t)}{T_{\rm in}} ,\
	v := \frac{pa^{1+q}(t)}{T_{\rm in}}\ ,
\end{align}
one can reexpress the series coefficients $f^{(0)}_k$'s as
\begin{align}
	f^{(0)}_k
		&= \left( -\alpha(1-q) \frac{\tau_R}{t_{\rm in}} \right)^k u^{-\frac{k}{\alpha(1-q)}} \left( - v^{1-\frac{1}{\alpha(1-q)}} \frac{\partial}{\partial v}  \right)^k e^{-v} \nonumber \\
		&= \delta_{k,0} + \left(\frac{\tau_R}{t_{\rm in}}\right)^k u^{-\frac{k}{\alpha(1-q)}} \sum^\infty_{m=0} v^{1+m+\frac{k}{\alpha(1-q)}}  \nonumber\\
			&\quad\quad\quad\quad \times \frac{(-1)^{m+1}\Gamma\left(k+(m+1)\alpha(1-q)\right)}{(m+1)!\Gamma\left((m+1)\alpha(1-q)\right)} \ .  \label{eq63}
\end{align}
It is clear that $f^{(0)}_k$ is factorially divergent at large $k$.  Using this expression, one may obtain the perturbative series for the energy density $E^{(0)}$, which is defined as the first-order moment of the distribution function $f^{(0)}$ in the kinetic theory as
\begin{align}\label{eq:pert_e_kine}
	E^{(0)}(t)
			  &:= \int \frac{d^3\vec{p}}{(2\pi)^3} \sqrt{-g(t)}pa(t) f^{(0)}(t,p) \nonumber\\
			  &=: \sum_{k=0}^\infty E_k^{(0)}(t) \epsilon^k\ ,
\end{align}
where the coefficients $E_k^{(0)}$'s read
\begin{align}
	E_k^{(0)}
		&= \int \frac{d^3\vec{p}}{(2\pi)^3} \sqrt{-g}pa f^{(0)}_k \nonumber\\
		&=\frac{3 T^4}{\pi^2}\left(\frac{t}{\tau_R} \right)^{-k} \frac{\Gamma\left(k-4\alpha(1-q)\right)}{\Gamma\left(-4\alpha(1-q)\right)}  \ . \label{eq74}
\end{align}
Thus, the energy density $E^{(0)}$ in the kinetic theory is also expanded as a factorially divergent series within the CE method.

One can cure the factorial divergence and obtain a hydrodynamic attractor with the Borel resummation technique.  The Borel transform of the series (\ref{eq74}) reads
\begin{align}
	\mathcal{B}[E^{(0)}](s)
		&= \frac{3T^4}{\pi^2} \left(1-s\frac{\tau_R}{t}\right)^{4\alpha(1-q)}\ . \label{eq75}
\end{align}
The Borel transform of $E^{(0)}$ has a branch cut on the real axis originating from $s=t/\tau_R$ to the infinity.  The difference between $+$- and $-$-lateral Borel resummations is
\begin{align}
	(\mathcal{S}_{+} - \mathcal{S}_{-})[E^{(0)}]
		&= \frac{3T^4}{\pi^2}(-2i) \sin(4\alpha(1-q)\pi) \nonumber\\
		&\quad \times \Gamma(1+4\alpha(1-q)) \nonumber\\
		&\quad \times \left(\frac{t}{\tau_R} \right)^{-4\alpha(1-q)} {\rm e}^{-t/\tau_R}
\ , \label{eq76}
\end{align}
where $\epsilon=1$ is understood.  One can read out from the last line of Eq.~(\ref{eq76}) that the perturbative expansion on the one-instanton background $E^{(1)}$ is just a single term, instead of an infinite sum.  This means that the trans-series expansion terminates at $n=1$, and thus there exists the first-order non-hydrodynamic mode only.  With the median resummation, one can cancel the imaginary ambiguity and obtain a hydrodynamic attractor solution for the energy density $E^{(0)}$ in the kinetic theory as
\begin{align}
	&\mathcal{S}_{\rm med}[E^{(0)}] \nonumber\\
			&=\frac{3T^4}{\pi^2} \frac{t}{\tau_R} \frac{e^{- \frac{t}{\tau_R}}}{1+4\alpha(1-q)} \nonumber\\
				&\quad \times \Biggl[ \left(\frac{t}{\tau_R}\right)^{-1-4\alpha(1-q)} \Gamma\left(2+4\alpha(1-q)\right) \cos\left( 4\alpha(1-q) \pi \right) \nonumber\\
				&\quad\quad + {}_1F_1\left( \begin{array}{c}1+4\alpha(1-q)\\ 2+4\alpha(1-q) \end{array}; \frac{t}{\tau_R}\right)  \Biggl]\ .\label{eq77}
\end{align}
The hydrodynamic attractor solution (\ref{eq77}) is globally attractive, as in the case of the MIS theory.  Indeed, the expressions for the general solution to $f$ (\ref{eq43}) and its moment $E$ (\ref{eq45}) imply that variations (with arbitrary size) from an attractor solution decay exponentially.  One can also confirm that the attractor solution (\ref{eq77}) includes the all-order gradients $(\tau_R/t)^k$ and behaves in the late-time limit as
\begin{align}\label{eq78}
	&\mathcal{S}_{\rm med }[E^{(0)}] \nonumber\\
		&= \frac{3T^4}{\pi^2} \left[ 1 - 4\alpha(1-q) \frac{\tau_R}{t} \right. \nonumber\\
		&\quad \left. - 4\alpha (1-q) \left( 1 -4\alpha(1-q) \right) \left( \frac{\tau_R}{t} \right)^2 +\mathcal{O}\left( \left( \frac{\tau_R}{t} \right)^3\right) \right] \ ,
\end{align}
which is in agreement with the perturbative series for the energy density (\ref{eq74}).

\begin{figure}
\includegraphics[width=0.45\textwidth]{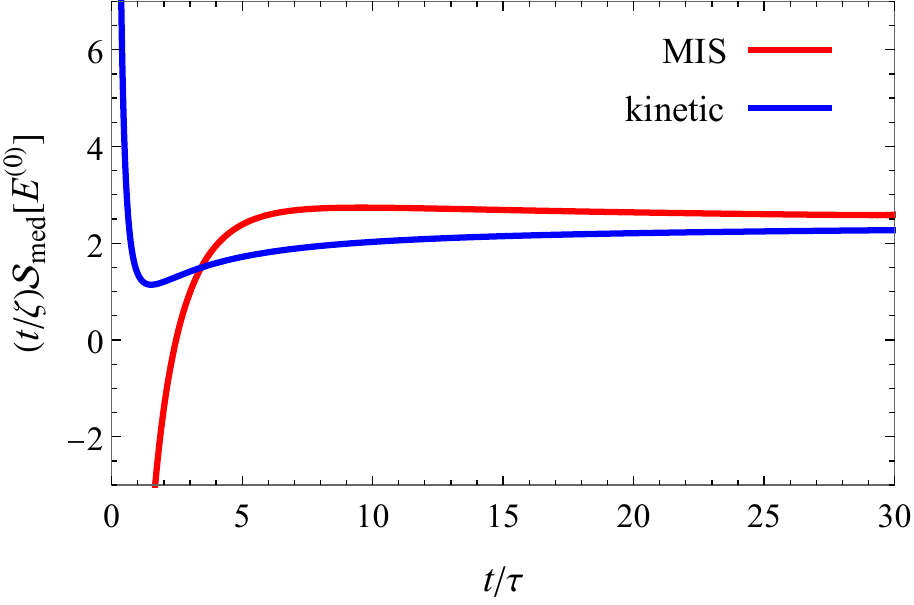}
\caption{A comparison between the hydrodynamic attractor solutions for the energy density in the MIS (\ref{eq:resumE}) (red) and kinetic (\ref{eq77}) (blue) theories.  The horizontal axis is scaled with $\tau = \tau_\Pi$ and $\tau_R$ for the MIS and kinetic theories, respectively.  We choose the parameters as $\alpha=2/3$ and $c_s^2=1/3$.}\label{fig:e_KT1}
\end{figure}

The hydrodynamic attractor solution in the kinetic theory (\ref{eq77}) disagrees with that in the MIS theory (\ref{eq:resumE}).  As demonstrated in Fig.~\ref{fig:e_KT1}, the disagreement becomes larger at early times, where the gradient becomes large and therefore different microscopic theories give different results.  The two theories coincide with each other only at the very late times, where we imposed the matching condition (\ref{kin:temp2}).  At the analytical level, the late time behaviors in both theories are already derived in Eqs.~(\ref{eq28}) and (\ref{eq78}), which indicate that only the leading-order terms coincide with each other and that all the other terms disagree due to non-equilibrium corrections.  For early-time asymptotic behaviors, we find
\begin{align}
	&{\mathcal S}_{\rm med}[E^{(0)}] \xrightarrow{t \to 0} \label{eq58}\\  
	& \left\{ \begin{array}{ll} \displaystyle 3\zeta \frac{\alpha}{t} \left( \frac{t}{\tau} \right)^{1-Q}\Gamma(Q) \cos(Q \pi) \times \frac{3\alpha \pi}{\sin(Q\pi)} & ({\rm MIS}) \\[8pt] \displaystyle 3\zeta \frac{\alpha}{t}  \left( \frac{t}{\tau} \right)^{1-4\alpha}\Gamma(4\alpha) \cos(4\alpha \pi) \times \frac{3\alpha}{1-Q} & ({\rm kinetic}) \end{array} \right. \nonumber ,
\end{align}
where we used the matching condition (\ref{eq78}).  Both theories predict divergent behaviors at $t 
\to 0$ because $Q>1$ and $\alpha>(3(c_s^2+1))^{-1}>0$ [see discussions below Eq.~(\ref{eq20})].  The exponents are in general distinct and coincide only when $Q=4\alpha¥ \Leftrightarrow ¥ c_s^2=1/3$.  Even when $Q=4\alpha$, the prefactors for the divergences do not agree because of the last factors in Eq.~(\ref{eq58}).  It is notable that the signs of the early-time behaviors can be opposite, which is actually the case for the parameter choice in Fig.~\ref{fig:e_KT1}, because they are controlled by independent factors, $\tan(Q \pi)$ and $-\cos(4\alpha\pi)$, for the MIS and kinetic theories respectively.

\begin{figure}
\includegraphics[width=0.45\textwidth]{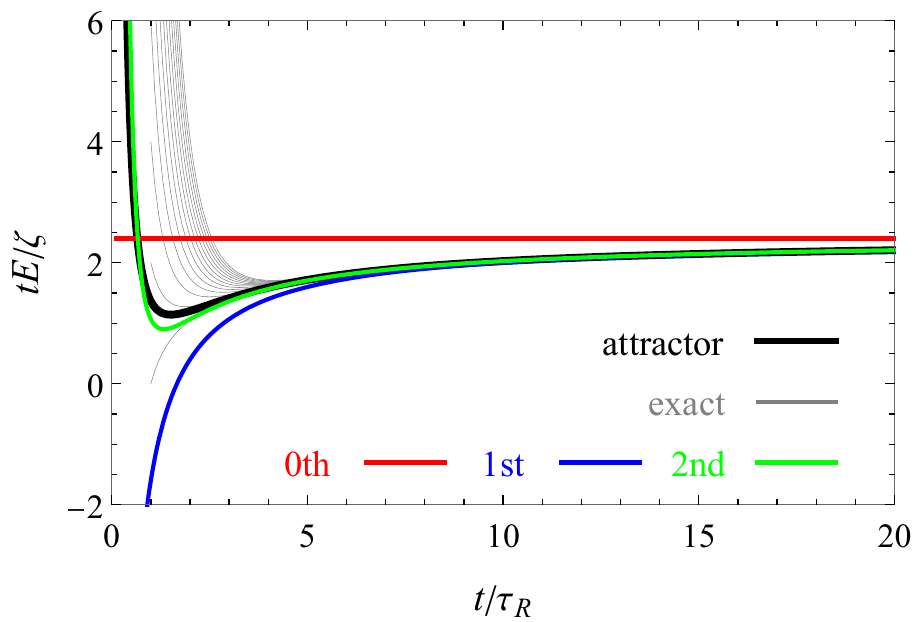}
\caption{The hydrodynamic attractor solution for the energy density in the kinetic theory (\ref{eq77}) (black line), in comparison with the exact solutions to the Boltzmann equation (\ref{eq45}) with various initial energies $E_{\rm in}$ (gray lines) and low-order results (\ref{eq74}) (red, blue, and green lines).  The parameters are set as $\alpha=2/3$ and $t_{\rm in}=\tau_R$, and the initial energies are set as $t_{\rm in} E_{\rm in}/\zeta = 2(m-1)\ (m=1,2 \cdots,14)$.}\label{fig:e_KT2}
\end{figure}

Figure~\ref{fig:e_KT2} shows the hydrodynamic attractor solution in the kinetic theory (\ref{eq77}), in comparisons with the energy density calculated from the exact solutions to Boltzmann equation (\ref{eq45}) and the low-order truncations of energy density in Eq.~(\ref{eq74}).  As in the case of the MIS theory (see Fig.~\ref{fig:att1}), no matter what the initial conditions are, the exact results approach quickly to the hydrodynamic attractor well before the low-order results work, with a typical time scale set by the relaxation time $\tau_R$.  The difference from the MIS theory is the speed of the convergence, i.e., non-hydrodynamics modes decay exponentially in the kinetic theory [see Eq.~(\ref{eq45})] whereas it is power in the MIS theory [see Eq.~(\ref{eq:23b})].  This is a fingerprint of the relaxation-type collision kernel in the Boltzmann equation (\ref{eq:Boltzmann}), for which the variation $\delta E$ satisfies a linear equation no matter how large they are.

\section{Summary} \label{sec:5}

We have analytically investigated the hydrodynamic attractor solutions in both MIS and kinetic theories for a viscous fluid system undergoing a manual Hubble expansion based on the Borel resummation technique.  To the best of our knowledge, this is the first study for hydrodynamic attractors in a Hubble flow.

In the MIS theory (see Sec.~\ref{sec:3}), we have shown that the gradient expansion (i.e., expansion in terms of $\tau_\Pi/t$ with $\tau_\Pi$ being relaxation time) leads to factorially divergent series for the bulk stress and energy density.  By resumming these divergent series, we have obtained the hydrodynamic attractor solutions.  We have compared the obtained hydrodynamic attractors with numerical solutions of the MIS theory, finding a good agreement well before the low-order hydrodynamics' work.  By examining the stability properties of the attractor solutions against initial conditions, we have shown that the hydrodynamic attractors in the Hubble flow are globally attractive, which is the same as those in the Bjorken flow but different from those in the Gubser flow due to its distinct dimension of the phase space \cite{Behtash18}.  The disagreements between the hydrodynamic attractors and the numerical solutions are caused by non-hydrodynamic modes, and we have found that only the first-order non-hydrodynamic mode $\propto e^{-t/\tau_\Pi}$ exists and all the higher-order contributions $\propto e^{-nt/\tau_\Pi}$ ($n\geq 2$) are vanishing.  We have also computed the Borel resummed effective bulk viscosity from the hydrodynamic attractor solution for the bulk stress and observed a strong suppression in large gradient region, similar to that for effective shear viscosity in the Bjorken flow shown in Ref.~\cite{romatschke_2}.

In the kinetic theory (see Sec.~\ref{sec:4}), by employing a relaxation-time approximation for the Boltzmann equation, we have shown that the CE expansion results in factorially divergent series for the distribution function and the energy density.  Applying the Borel resummation, we have obtained the hydrodynamic attractor solution for the energy density in the kinetic theory, which has been compared with exact solutions to the Boltzmann equation and found a good agreement well before the low-order truncations of the CE expansion work.  We have shown that the attractor is globally attractive and that the higher-order non-hydrodynamic modes $n\geq 2$ are absent, which are similar to the MIS theory. In contrast to the MIS theory, the speed of the attraction is exponential in the kinetic theory, while that is power in the MIS theory, due to the relaxation-type linear collision term for the Boltzmann equation.  We have also found that the early-time behaviors of the two attractors are quite different because of non-equilibrium corrections stemming from large gradients.

As for future work, it is interesting to extend our analysis to more realistic physical setups.  For example, in the present paper we have focused on the flat Hubble flow under a manually controlled expansion rate.  We have also assumed that the equation of state  takes a simple form and also the transport coefficients are constants.  Those simplifications have enabled us to perform all the calculations fully analytically, but are not so realistic when considering to apply to, e.g., actual cosmological problems, where the expansion rate should be determined self-consistently by solving the Einstein equation rather than being fixed.  Relaxing those simplifications would affect the hydrodynamic attractors and the associated non-hydrodynamic modes by modifying the structures of the factorial divergences as well as the resulting Borel transform/resummation, which can be studied, e.g., numerically based on the Pad\'{e}-Borel approximation and/or analytically by considering perturbations around the present results.

\section*{Acknowledgments}
We acknowledge Jorge~Noronha and Li~Yan for useful discussions.  We also thank Yukawa Institute for Theoretical Physics at Kyoto University.   Discussions during the YITP-RIKEN iTHEMS workshop YITP-T-20-03 on "Potential Toolkit to Attack Nonperturbative Aspects of QFT -Resurgence and related topics-" were useful to complete this work.  Z.~D and X.~G.~H are supported by NSFC through Grant No.~12075061 and Shanghai NSF through Grant No.~20ZR1404100.  

\bibliographystyle{apsrev4-2}
\bibliography{ref}

\begin{thebibliography}{68}%
\makeatletter
\providecommand \@ifxundefined [1]{%
 \@ifx{#1\undefined}
}%
\providecommand \@ifnum [1]{%
 \ifnum #1\expandafter \@firstoftwo
 \else \expandafter \@secondoftwo
 \fi
}%
\providecommand \@ifx [1]{%
 \ifx #1\expandafter \@firstoftwo
 \else \expandafter \@secondoftwo
 \fi
}%
\providecommand \natexlab [1]{#1}%
\providecommand \enquote  [1]{``#1''}%
\providecommand \bibnamefont  [1]{#1}%
\providecommand \bibfnamefont [1]{#1}%
\providecommand \citenamefont [1]{#1}%
\providecommand \href@noop [0]{\@secondoftwo}%
\providecommand \href [0]{\begingroup \@sanitize@url \@href}%
\providecommand \@href[1]{\@@startlink{#1}\@@href}%
\providecommand \@@href[1]{\endgroup#1\@@endlink}%
\providecommand \@sanitize@url [0]{\catcode `\\12\catcode `\$12\catcode
  `\&12\catcode `\#12\catcode `\^12\catcode `\_12\catcode `\%12\relax}%
\providecommand \@@startlink[1]{}%
\providecommand \@@endlink[0]{}%
\providecommand \url  [0]{\begingroup\@sanitize@url \@url }%
\providecommand \@url [1]{\endgroup\@href {#1}{\urlprefix }}%
\providecommand \urlprefix  [0]{URL }%
\providecommand \Eprint [0]{\href }%
\providecommand \doibase [0]{https://doi.org/}%
\providecommand \selectlanguage [0]{\@gobble}%
\providecommand \bibinfo  [0]{\@secondoftwo}%
\providecommand \bibfield  [0]{\@secondoftwo}%
\providecommand \translation [1]{[#1]}%
\providecommand \BibitemOpen [0]{}%
\providecommand \bibitemStop [0]{}%
\providecommand \bibitemNoStop [0]{.\EOS\space}%
\providecommand \EOS [0]{\spacefactor3000\relax}%
\providecommand \BibitemShut  [1]{\csname bibitem#1\endcsname}%
\let\auto@bib@innerbib\@empty
\bibitem [{\citenamefont {Weinberg}(1972)}]{Weinberg:1972kfs}%
  \BibitemOpen
  \bibfield  {author} {\bibinfo {author} {\bibfnamefont {S.}~\bibnamefont
  {Weinberg}},\ }\href@noop {} {\emph {\bibinfo {title} {{Gravitation and
  Cosmology}: {Principles and Applications of the General Theory of
  Relativity}}}}\ (\bibinfo  {publisher} {John Wiley and Sons},\ \bibinfo
  {address} {New York},\ \bibinfo {year} {1972})\BibitemShut {NoStop}%
\bibitem [{\citenamefont {Brevik}\ \emph {et~al.}(2017)\citenamefont {Brevik},
  \citenamefont {Gr{\o}n}, \citenamefont {de~Haro}, \citenamefont {Odintsov},\
  and\ \citenamefont {Saridakis}}]{Brevik_2017}%
  \BibitemOpen
  \bibfield  {author} {\bibinfo {author} {\bibfnamefont {I.}~\bibnamefont
  {Brevik}}, \bibinfo {author} {\bibfnamefont {{\O}.}~\bibnamefont {Gr{\o}n}},
  \bibinfo {author} {\bibfnamefont {J.}~\bibnamefont {de~Haro}}, \bibinfo
  {author} {\bibfnamefont {S.~D.}\ \bibnamefont {Odintsov}},\ and\ \bibinfo
  {author} {\bibfnamefont {E.~N.}\ \bibnamefont {Saridakis}},\ }\href
  {https://doi.org/10.1142/s0218271817300245} {\bibfield  {journal} {\bibinfo
  {journal} {Int. J. Mod. Phys. D}\ }\textbf {\bibinfo {volume} {26}},\
  \bibinfo {pages} {1730024} (\bibinfo {year} {2017})},\ \Eprint
  {https://arxiv.org/abs/1706.02543} {arXiv:1706.02543 [gr-qc]} \BibitemShut
  {NoStop}%
\bibitem [{\citenamefont {Rezzolla}\ and\ \citenamefont
  {Zanotti}(2013)}]{Rezzolla}%
  \BibitemOpen
  \bibfield  {author} {\bibinfo {author} {\bibfnamefont {L.}~\bibnamefont
  {Rezzolla}}\ and\ \bibinfo {author} {\bibfnamefont {O.}~\bibnamefont
  {Zanotti}},\ }\href@noop {} {\emph {\bibinfo {title} {{Relativistic
  Hydrodynamics}}}}\ (\bibinfo  {publisher} {Oxford University Press},\
  \bibinfo {address} {Oxford},\ \bibinfo {year} {2013})\BibitemShut {NoStop}%
\bibitem [{\citenamefont {Heinz}\ and\ \citenamefont
  {Snellings}(2013)}]{Heinz:2013th}%
  \BibitemOpen
  \bibfield  {author} {\bibinfo {author} {\bibfnamefont {U.}~\bibnamefont
  {Heinz}}\ and\ \bibinfo {author} {\bibfnamefont {R.}~\bibnamefont
  {Snellings}},\ }\href {https://doi.org/10.1146/annurev-nucl-102212-170540}
  {\bibfield  {journal} {\bibinfo  {journal} {Ann. Rev. Nucl. Part. Sci.}\
  }\textbf {\bibinfo {volume} {63}},\ \bibinfo {pages} {123} (\bibinfo {year}
  {2013})},\ \Eprint {https://arxiv.org/abs/1301.2826} {arXiv:1301.2826
  [nucl-th]} \BibitemShut {NoStop}%
\bibitem [{\citenamefont {Gale}\ \emph {et~al.}(2013)\citenamefont {Gale},
  \citenamefont {Jeon},\ and\ \citenamefont {Schenke}}]{Gale:2013da}%
  \BibitemOpen
  \bibfield  {author} {\bibinfo {author} {\bibfnamefont {C.}~\bibnamefont
  {Gale}}, \bibinfo {author} {\bibfnamefont {S.}~\bibnamefont {Jeon}},\ and\
  \bibinfo {author} {\bibfnamefont {B.}~\bibnamefont {Schenke}},\ }\href
  {https://doi.org/10.1142/S0217751X13400113} {\bibfield  {journal} {\bibinfo
  {journal} {Int. J. Mod. Phys. A}\ }\textbf {\bibinfo {volume} {28}},\
  \bibinfo {pages} {1340011} (\bibinfo {year} {2013})},\ \Eprint
  {https://arxiv.org/abs/1301.5893} {arXiv:1301.5893 [nucl-th]} \BibitemShut
  {NoStop}%
\bibitem [{\citenamefont {Romatschke}\ and\ \citenamefont
  {Romatschke}(2019)}]{Romatschke:2017ejr}%
  \BibitemOpen
  \bibfield  {author} {\bibinfo {author} {\bibfnamefont {P.}~\bibnamefont
  {Romatschke}}\ and\ \bibinfo {author} {\bibfnamefont {U.}~\bibnamefont
  {Romatschke}},\ }\href@noop {} {\emph {\bibinfo {title} {{Relativistic Fluid
  Dynamics In and Out of Equilibrium}}}}\ (\bibinfo  {publisher} {Cambridge
  University Press},\ \bibinfo {address} {Cambridge},\ \bibinfo {year}
  {2019})\BibitemShut {NoStop}%
\bibitem [{\citenamefont {Florkowski}\ \emph {et~al.}(2018)\citenamefont
  {Florkowski}, \citenamefont {Heller},\ and\ \citenamefont
  {Spali{\'n}ski}}]{Florkowski_2018}%
  \BibitemOpen
  \bibfield  {author} {\bibinfo {author} {\bibfnamefont {W.}~\bibnamefont
  {Florkowski}}, \bibinfo {author} {\bibfnamefont {M.~P.}\ \bibnamefont
  {Heller}},\ and\ \bibinfo {author} {\bibfnamefont {M.}~\bibnamefont
  {Spali{\'n}ski}},\ }\href {https://doi.org/10.1088/1361-6633/aaa091}
  {\bibfield  {journal} {\bibinfo  {journal} {Rep. Prog. Phys.}\ }\textbf
  {\bibinfo {volume} {81}},\ \bibinfo {pages} {046001} (\bibinfo {year}
  {2018})},\ \Eprint {https://arxiv.org/abs/1707.02282} {arXiv:1707.02282
  [hep-ph]} \BibitemShut {NoStop}%
\bibitem [{\citenamefont {Shen}\ and\ \citenamefont {Yan}(2020)}]{Shen2020}%
  \BibitemOpen
  \bibfield  {author} {\bibinfo {author} {\bibfnamefont {C.}~\bibnamefont
  {Shen}}\ and\ \bibinfo {author} {\bibfnamefont {L.}~\bibnamefont {Yan}},\
  }\href {https://doi.org/10.1007/s41365-020-00829-z} {\bibfield  {journal}
  {\bibinfo  {journal} {Nucl. Sci. Tech.}\ }\textbf {\bibinfo {volume} {31}},\
  \bibinfo {pages} {122} (\bibinfo {year} {2020})},\ \Eprint
  {https://arxiv.org/abs/2010.12377} {arXiv:2010.12377 [nucl-th]} \BibitemShut
  {NoStop}%
\bibitem [{\citenamefont {Hiscock}\ and\ \citenamefont
  {Lindblom}(1983)}]{Hiscock:1983zz}%
  \BibitemOpen
  \bibfield  {author} {\bibinfo {author} {\bibfnamefont {W.~A.}\ \bibnamefont
  {Hiscock}}\ and\ \bibinfo {author} {\bibfnamefont {L.}~\bibnamefont
  {Lindblom}},\ }\href {https://doi.org/10.1016/0003-4916(83)90288-9}
  {\bibfield  {journal} {\bibinfo  {journal} {Annals Phys.}\ }\textbf {\bibinfo
  {volume} {151}},\ \bibinfo {pages} {466} (\bibinfo {year}
  {1983})}\BibitemShut {NoStop}%
\bibitem [{\citenamefont {Pu}\ \emph {et~al.}(2010)\citenamefont {Pu},
  \citenamefont {Koide},\ and\ \citenamefont {Rischke}}]{Pu:2009fj}%
  \BibitemOpen
  \bibfield  {author} {\bibinfo {author} {\bibfnamefont {S.}~\bibnamefont
  {Pu}}, \bibinfo {author} {\bibfnamefont {T.}~\bibnamefont {Koide}},\ and\
  \bibinfo {author} {\bibfnamefont {D.~H.}\ \bibnamefont {Rischke}},\ }\href
  {https://doi.org/10.1103/PhysRevD.81.114039} {\bibfield  {journal} {\bibinfo
  {journal} {Phys. Rev. D}\ }\textbf {\bibinfo {volume} {81}},\ \bibinfo
  {pages} {114039} (\bibinfo {year} {2010})},\ \Eprint
  {https://arxiv.org/abs/0907.3906} {arXiv:0907.3906 [hep-ph]} \BibitemShut
  {NoStop}%
\bibitem [{\citenamefont {M{\"u}ller}(1967)}]{Muller:1967zza}%
  \BibitemOpen
  \bibfield  {author} {\bibinfo {author} {\bibfnamefont {I.}~\bibnamefont
  {M{\"u}ller}},\ }\href {https://doi.org/10.1007/BF01326412} {\bibfield
  {journal} {\bibinfo  {journal} {Z. Phys.}\ }\textbf {\bibinfo {volume}
  {198}},\ \bibinfo {pages} {329} (\bibinfo {year} {1967})}\BibitemShut
  {NoStop}%
\bibitem [{\citenamefont {Israel}\ and\ \citenamefont
  {Stewart}(1979)}]{ISRAEL1979341}%
  \BibitemOpen
  \bibfield  {author} {\bibinfo {author} {\bibfnamefont {W.}~\bibnamefont
  {Israel}}\ and\ \bibinfo {author} {\bibfnamefont {J.}~\bibnamefont
  {Stewart}},\ }\href {https://doi.org/10.1016/0003-4916(79)90130-1} {\bibfield
   {journal} {\bibinfo  {journal} {Annals Phys.}\ }\textbf {\bibinfo {volume}
  {118}},\ \bibinfo {pages} {341} (\bibinfo {year} {1979})}\BibitemShut
  {NoStop}%
\bibitem [{\citenamefont {Weller}\ and\ \citenamefont
  {Romatschke}(2017)}]{Weller:2017tsr}%
  \BibitemOpen
  \bibfield  {author} {\bibinfo {author} {\bibfnamefont {R.~D.}\ \bibnamefont
  {Weller}}\ and\ \bibinfo {author} {\bibfnamefont {P.}~\bibnamefont
  {Romatschke}},\ }\href {https://doi.org/10.1016/j.physletb.2017.09.077}
  {\bibfield  {journal} {\bibinfo  {journal} {Phys. Lett. B}\ }\textbf
  {\bibinfo {volume} {774}},\ \bibinfo {pages} {351} (\bibinfo {year}
  {2017})},\ \Eprint {https://arxiv.org/abs/1701.07145} {arXiv:1701.07145
  [nucl-th]} \BibitemShut {NoStop}%
\bibitem [{\citenamefont {M{\"a}ntysaari}\ \emph {et~al.}(2017)\citenamefont
  {M{\"a}ntysaari}, \citenamefont {Schenke}, \citenamefont {Shen},\ and\
  \citenamefont {Tribedy}}]{Mantysaari:2017cni}%
  \BibitemOpen
  \bibfield  {author} {\bibinfo {author} {\bibfnamefont {H.}~\bibnamefont
  {M{\"a}ntysaari}}, \bibinfo {author} {\bibfnamefont {B.}~\bibnamefont
  {Schenke}}, \bibinfo {author} {\bibfnamefont {C.}~\bibnamefont {Shen}},\ and\
  \bibinfo {author} {\bibfnamefont {P.}~\bibnamefont {Tribedy}},\ }\href
  {https://doi.org/10.1016/j.physletb.2017.07.038} {\bibfield  {journal}
  {\bibinfo  {journal} {Phys. Lett. B}\ }\textbf {\bibinfo {volume} {772}},\
  \bibinfo {pages} {681} (\bibinfo {year} {2017})},\ \Eprint
  {https://arxiv.org/abs/1705.03177} {arXiv:1705.03177 [nucl-th]} \BibitemShut
  {NoStop}%
\bibitem [{\citenamefont {Habich}\ \emph {et~al.}(2016)\citenamefont {Habich},
  \citenamefont {Miller}, \citenamefont {Romatschke},\ and\ \citenamefont
  {Xiang}}]{Habich_2016}%
  \BibitemOpen
  \bibfield  {author} {\bibinfo {author} {\bibfnamefont {M.}~\bibnamefont
  {Habich}}, \bibinfo {author} {\bibfnamefont {G.~A.}\ \bibnamefont {Miller}},
  \bibinfo {author} {\bibfnamefont {P.}~\bibnamefont {Romatschke}},\ and\
  \bibinfo {author} {\bibfnamefont {W.}~\bibnamefont {Xiang}},\ }\href
  {https://doi.org/10.1140/epjc/s10052-016-4237-z} {\bibfield  {journal}
  {\bibinfo  {journal} {Eur. Phys. J. C}\ }\textbf {\bibinfo {volume} {76}},\
  \bibinfo {pages} {408} (\bibinfo {year} {2016})},\ \Eprint
  {https://arxiv.org/abs/1512.05354} {arXiv:1512.05354 [nucl-th]} \BibitemShut
  {NoStop}%
\bibitem [{\citenamefont {Nagle}\ \emph {et~al.}(2014)\citenamefont {Nagle},
  \citenamefont {Adare}, \citenamefont {Beckman}, \citenamefont {Koblesky},
  \citenamefont {{Orjuela Koop}}, \citenamefont {McGlinchey}, \citenamefont
  {Romatschke}, \citenamefont {Carlson}, \citenamefont {Lynn},\ and\
  \citenamefont {McCumber}}]{Nagle:2013lja}%
  \BibitemOpen
  \bibfield  {author} {\bibinfo {author} {\bibfnamefont {J.~L.}\ \bibnamefont
  {Nagle}}, \bibinfo {author} {\bibfnamefont {A.}~\bibnamefont {Adare}},
  \bibinfo {author} {\bibfnamefont {S.}~\bibnamefont {Beckman}}, \bibinfo
  {author} {\bibfnamefont {T.}~\bibnamefont {Koblesky}}, \bibinfo {author}
  {\bibfnamefont {J.}~\bibnamefont {{Orjuela Koop}}}, \bibinfo {author}
  {\bibfnamefont {D.}~\bibnamefont {McGlinchey}}, \bibinfo {author}
  {\bibfnamefont {P.}~\bibnamefont {Romatschke}}, \bibinfo {author}
  {\bibfnamefont {J.}~\bibnamefont {Carlson}}, \bibinfo {author} {\bibfnamefont
  {J.~E.}\ \bibnamefont {Lynn}},\ and\ \bibinfo {author} {\bibfnamefont
  {M.}~\bibnamefont {McCumber}},\ }\href
  {https://doi.org/10.1103/PhysRevLett.113.112301} {\bibfield  {journal}
  {\bibinfo  {journal} {Phys. Rev. Lett.}\ }\textbf {\bibinfo {volume} {113}},\
  \bibinfo {pages} {112301} (\bibinfo {year} {2014})},\ \Eprint
  {https://arxiv.org/abs/1312.4565} {arXiv:1312.4565 [nucl-th]} \BibitemShut
  {NoStop}%
\bibitem [{\citenamefont {Heller}\ \emph {et~al.}(2013)\citenamefont {Heller},
  \citenamefont {Janik},\ and\ \citenamefont {Witaszczyk}}]{Heller_2013}%
  \BibitemOpen
  \bibfield  {author} {\bibinfo {author} {\bibfnamefont {M.~P.}\ \bibnamefont
  {Heller}}, \bibinfo {author} {\bibfnamefont {R.~A.}\ \bibnamefont {Janik}},\
  and\ \bibinfo {author} {\bibfnamefont {P.}~\bibnamefont {Witaszczyk}},\
  }\href {https://doi.org/10.1103/physrevlett.110.211602} {\bibfield  {journal}
  {\bibinfo  {journal} {Phys. Rev. Lett.}\ }\textbf {\bibinfo {volume} {110}},\
  \bibinfo {pages} {211602} (\bibinfo {year} {2013})},\ \Eprint
  {https://arxiv.org/abs/1302.0697} {arXiv:1302.0697 [hep-th]} \BibitemShut
  {NoStop}%
\bibitem [{\citenamefont {Heller}\ and\ \citenamefont
  {Spali{\'n}ski}(2015)}]{Heller_2015}%
  \BibitemOpen
  \bibfield  {author} {\bibinfo {author} {\bibfnamefont {M.~P.}\ \bibnamefont
  {Heller}}\ and\ \bibinfo {author} {\bibfnamefont {M.}~\bibnamefont
  {Spali{\'n}ski}},\ }\href {https://doi.org/10.1103/physrevlett.115.072501}
  {\bibfield  {journal} {\bibinfo  {journal} {Phys. Rev. Lett.}\ }\textbf
  {\bibinfo {volume} {115}},\ \bibinfo {pages} {072501} (\bibinfo {year}
  {2015})},\ \Eprint {https://arxiv.org/abs/1503.07514} {arXiv:1503.07514
  [hep-th]} \BibitemShut {NoStop}%
\bibitem [{\citenamefont {Ecalle}(1981)}]{Ecalle}%
  \BibitemOpen
  \bibfield  {author} {\bibinfo {author} {\bibfnamefont {J.}~\bibnamefont
  {Ecalle}},\ }\href@noop {} {\emph {\bibinfo {title} {Les Fonctions
  Resurgentes}}},\ Vol.\ \bibinfo {volume} {I-III}\ (\bibinfo  {publisher}
  {Publications Math{\'e}matiques d'Orsay},\ \bibinfo {year}
  {1981})\BibitemShut {NoStop}%
\bibitem [{\citenamefont {Aniceto}\ \emph {et~al.}(2019)\citenamefont
  {Aniceto}, \citenamefont {Ba{\c s}ar},\ and\ \citenamefont
  {Schiappa}}]{ANICETO20191}%
  \BibitemOpen
  \bibfield  {author} {\bibinfo {author} {\bibfnamefont {I.}~\bibnamefont
  {Aniceto}}, \bibinfo {author} {\bibfnamefont {G.}~\bibnamefont {Ba{\c
  s}ar}},\ and\ \bibinfo {author} {\bibfnamefont {R.}~\bibnamefont
  {Schiappa}},\ }\href {https://doi.org/10.1016/j.physrep.2019.02.003}
  {\bibfield  {journal} {\bibinfo  {journal} {Phys. Rept.}\ }\textbf {\bibinfo
  {volume} {809}},\ \bibinfo {pages} {1} (\bibinfo {year} {2019})},\ \Eprint
  {https://arxiv.org/abs/1802.10441} {arXiv:1802.10441 [hep-th]} \BibitemShut
  {NoStop}%
\bibitem [{\citenamefont {Aniceto}\ and\ \citenamefont {{Spali\ifmmode
  \acute{n}\else {\'n}\fi{}ski}}(2016)}]{Aniceto2016}%
  \BibitemOpen
  \bibfield  {author} {\bibinfo {author} {\bibfnamefont {I.}~\bibnamefont
  {Aniceto}}\ and\ \bibinfo {author} {\bibfnamefont {M.}~\bibnamefont
  {{Spali\ifmmode \acute{n}\else {\'n}\fi{}ski}}},\ }\href
  {https://doi.org/10.1103/PhysRevD.93.085008} {\bibfield  {journal} {\bibinfo
  {journal} {Phys. Rev. D}\ }\textbf {\bibinfo {volume} {93}},\ \bibinfo
  {pages} {085008} (\bibinfo {year} {2016})},\ \Eprint
  {https://arxiv.org/abs/1511.06358} {arXiv:1511.06358 [hep-th]} \BibitemShut
  {NoStop}%
\bibitem [{\citenamefont {Ba{\c s}ar}\ and\ \citenamefont
  {Dunne}(2015)}]{Ba_ar_2015}%
  \BibitemOpen
  \bibfield  {author} {\bibinfo {author} {\bibfnamefont {G.}~\bibnamefont
  {Ba{\c s}ar}}\ and\ \bibinfo {author} {\bibfnamefont {G.~V.}\ \bibnamefont
  {Dunne}},\ }\href {https://doi.org/10.1103/physrevd.92.125011} {\bibfield
  {journal} {\bibinfo  {journal} {Phys. Rev. D}\ }\textbf {\bibinfo {volume}
  {92}},\ \bibinfo {pages} {125011} (\bibinfo {year} {2015})},\ \Eprint
  {https://arxiv.org/abs/1509.05046} {arXiv:1509.05046 [hep-th]} \BibitemShut
  {NoStop}%
\bibitem [{\citenamefont {Heller}\ \emph {et~al.}(2018)\citenamefont {Heller},
  \citenamefont {Kurkela}, \citenamefont {{Spali\ifmmode \acute{n}\else
  {\'n}\fi{}ski}},\ and\ \citenamefont {Svensson}}]{PhysRevD.97.091503}%
  \BibitemOpen
  \bibfield  {author} {\bibinfo {author} {\bibfnamefont {M.~P.}\ \bibnamefont
  {Heller}}, \bibinfo {author} {\bibfnamefont {A.}~\bibnamefont {Kurkela}},
  \bibinfo {author} {\bibfnamefont {M.}~\bibnamefont {{Spali\ifmmode
  \acute{n}\else {\'n}\fi{}ski}}},\ and\ \bibinfo {author} {\bibfnamefont
  {V.}~\bibnamefont {Svensson}},\ }\href
  {https://doi.org/10.1103/PhysRevD.97.091503} {\bibfield  {journal} {\bibinfo
  {journal} {Phys. Rev. D}\ }\textbf {\bibinfo {volume} {97}},\ \bibinfo
  {pages} {091503} (\bibinfo {year} {2018})},\ \Eprint
  {https://arxiv.org/abs/1609.04803} {arXiv:1609.04803 [nucl-th]} \BibitemShut
  {NoStop}%
\bibitem [{\citenamefont {Denicol}\ and\ \citenamefont
  {Noronha}(2017)}]{denicol2017analytical}%
  \BibitemOpen
  \bibfield  {author} {\bibinfo {author} {\bibfnamefont {G.~S.}\ \bibnamefont
  {Denicol}}\ and\ \bibinfo {author} {\bibfnamefont {J.}~\bibnamefont
  {Noronha}},\ }\href {https://doi.org/10.1103/PhysRevD.97.056021} {\bibfield
  {journal} {\bibinfo  {journal} {Phys. Rev. D}\ }\textbf {\bibinfo {volume}
  {97}},\ \bibinfo {pages} {056021} (\bibinfo {year} {2017})},\ \Eprint
  {https://arxiv.org/abs/1711.01657} {arXiv:1711.01657 [nucl-th]} \BibitemShut
  {NoStop}%
\bibitem [{\citenamefont {Romatschke}(2018)}]{romatschke_2}%
  \BibitemOpen
  \bibfield  {author} {\bibinfo {author} {\bibfnamefont {P.}~\bibnamefont
  {Romatschke}},\ }\href {https://doi.org/10.1103/PhysRevLett.120.012301}
  {\bibfield  {journal} {\bibinfo  {journal} {Phys. Rev. Lett.}\ }\textbf
  {\bibinfo {volume} {120}},\ \bibinfo {pages} {012301} (\bibinfo {year}
  {2018})},\ \Eprint {https://arxiv.org/abs/1704.08699} {arXiv:1704.08699
  [hep-th]} \BibitemShut {NoStop}%
\bibitem [{\citenamefont {Blaizot}\ and\ \citenamefont
  {Yan}(2017)}]{Blaizot:2017lht}%
  \BibitemOpen
  \bibfield  {author} {\bibinfo {author} {\bibfnamefont {J.-P.}\ \bibnamefont
  {Blaizot}}\ and\ \bibinfo {author} {\bibfnamefont {L.}~\bibnamefont {Yan}},\
  }\href {https://doi.org/10.1007/JHEP11(2017)161} {\bibfield  {journal}
  {\bibinfo  {journal} {J. High Energy Phys.}\ }\textbf {\bibinfo {volume}
  {11}},\ \bibinfo {pages} {161}},\ \Eprint {https://arxiv.org/abs/1703.10634}
  {arXiv:1703.10634 [nucl-th]} \BibitemShut {NoStop}%
\bibitem [{\citenamefont {Behtash}\ \emph {et~al.}(2018)\citenamefont
  {Behtash}, \citenamefont {Cruz-Camacho},\ and\ \citenamefont
  {Martinez}}]{Behtash18}%
  \BibitemOpen
  \bibfield  {author} {\bibinfo {author} {\bibfnamefont {A.}~\bibnamefont
  {Behtash}}, \bibinfo {author} {\bibfnamefont {C.~N.}\ \bibnamefont
  {Cruz-Camacho}},\ and\ \bibinfo {author} {\bibfnamefont {M.}~\bibnamefont
  {Martinez}},\ }\href {https://doi.org/10.1103/PhysRevD.97.044041} {\bibfield
  {journal} {\bibinfo  {journal} {Phys. Rev. D}\ }\textbf {\bibinfo {volume}
  {97}},\ \bibinfo {pages} {044041} (\bibinfo {year} {2018})},\ \Eprint
  {https://arxiv.org/abs/1711.01745} {arXiv:1711.01745 [hep-th]} \BibitemShut
  {NoStop}%
\bibitem [{\citenamefont {Denicol}\ and\ \citenamefont
  {Noronha}(2019)}]{Denicol_2019}%
  \BibitemOpen
  \bibfield  {author} {\bibinfo {author} {\bibfnamefont {G.~S.}\ \bibnamefont
  {Denicol}}\ and\ \bibinfo {author} {\bibfnamefont {J.}~\bibnamefont
  {Noronha}},\ }\href {https://doi.org/10.1103/physrevd.99.116004} {\bibfield
  {journal} {\bibinfo  {journal} {Phys. Rev. D}\ }\textbf {\bibinfo {volume}
  {99}},\ \bibinfo {pages} {116004} (\bibinfo {year} {2019})},\ \Eprint
  {https://arxiv.org/abs/1804.04771} {arXiv:1804.04771 [nucl-th]} \BibitemShut
  {NoStop}%
\bibitem [{\citenamefont {Zinn-Justin}(1981)}]{Zinn-Justin:1981:PsloqmaftApr}%
  \BibitemOpen
  \bibfield  {author} {\bibinfo {author} {\bibfnamefont {J.}~\bibnamefont
  {Zinn-Justin}},\ }\href {https://doi.org/10.1016/0370-1573(81)90016-8}
  {\bibfield  {journal} {\bibinfo  {journal} {Phys. Rept.}\ }\textbf {\bibinfo
  {volume} {70}},\ \bibinfo {pages} {109} (\bibinfo {year} {1981})}\BibitemShut
  {NoStop}%
\bibitem [{\citenamefont {Jentschura}\ and\ \citenamefont
  {Zinn-Justin}(2004)}]{Jentschura_2004}%
  \BibitemOpen
  \bibfield  {author} {\bibinfo {author} {\bibfnamefont {U.~D.}\ \bibnamefont
  {Jentschura}}\ and\ \bibinfo {author} {\bibfnamefont {J.}~\bibnamefont
  {Zinn-Justin}},\ }\href {https://doi.org/10.1016/j.physletb.2004.06.077}
  {\bibfield  {journal} {\bibinfo  {journal} {Phys. Lett. B}\ }\textbf
  {\bibinfo {volume} {596}},\ \bibinfo {pages} {138} (\bibinfo {year}
  {2004})},\ \Eprint {https://arxiv.org/abs/hep-ph/0405279}
  {arXiv:hep-ph/0405279} \BibitemShut {NoStop}%
\bibitem [{\citenamefont {Misumi}\ \emph {et~al.}(2015)\citenamefont {Misumi},
  \citenamefont {Nitta},\ and\ \citenamefont {Sakai}}]{Misumi_2015}%
  \BibitemOpen
  \bibfield  {author} {\bibinfo {author} {\bibfnamefont {T.}~\bibnamefont
  {Misumi}}, \bibinfo {author} {\bibfnamefont {M.}~\bibnamefont {Nitta}},\ and\
  \bibinfo {author} {\bibfnamefont {N.}~\bibnamefont {Sakai}},\ }\href
  {https://doi.org/10.1007/jhep09(2015)157} {\bibfield  {journal} {\bibinfo
  {journal} {J. High Energy Phys.}\ }\textbf {\bibinfo {volume} {09}},\
  \bibinfo {pages} {157}},\ \Eprint {https://arxiv.org/abs/1507.00408}
  {arXiv:1507.00408 [hep-th]} \BibitemShut {NoStop}%
\bibitem [{\citenamefont {Dunne}\ and\ \citenamefont
  {{\"U}nsal}(2012)}]{Dunne_2012}%
  \BibitemOpen
  \bibfield  {author} {\bibinfo {author} {\bibfnamefont {G.~V.}\ \bibnamefont
  {Dunne}}\ and\ \bibinfo {author} {\bibfnamefont {M.}~\bibnamefont
  {{\"U}nsal}},\ }\href {https://doi.org/10.1007/jhep11(2012)170} {\bibfield
  {journal} {\bibinfo  {journal} {J. High Energy Phys.}\ }\textbf {\bibinfo
  {volume} {11}},\ \bibinfo {pages} {170}},\ \Eprint
  {https://arxiv.org/abs/1210.2423} {arXiv:1210.2423 [hep-th]} \BibitemShut
  {NoStop}%
\bibitem [{\citenamefont {Aniceto}\ and\ \citenamefont
  {Schiappa}(2015)}]{AnicSchi15:NonpeAmbigandRealiResurTrans}%
  \BibitemOpen
  \bibfield  {author} {\bibinfo {author} {\bibfnamefont {I.}~\bibnamefont
  {Aniceto}}\ and\ \bibinfo {author} {\bibfnamefont {R.}~\bibnamefont
  {Schiappa}},\ }\href {https://doi.org/10.1007/s00220-014-2165-z} {\bibfield
  {journal} {\bibinfo  {journal} {Commun. Math. Phys.}\ }\textbf {\bibinfo
  {volume} {335}},\ \bibinfo {pages} {183} (\bibinfo {year} {2015})},\ \Eprint
  {https://arxiv.org/abs/1308.1115} {arXiv:1308.1115 [hep-th]} \BibitemShut
  {NoStop}%
\bibitem [{\citenamefont {Pasquetti}\ and\ \citenamefont
  {Schiappa}(2010)}]{Pasquetti_2010}%
  \BibitemOpen
  \bibfield  {author} {\bibinfo {author} {\bibfnamefont {S.}~\bibnamefont
  {Pasquetti}}\ and\ \bibinfo {author} {\bibfnamefont {R.}~\bibnamefont
  {Schiappa}},\ }\href {https://doi.org/10.1007/s00023-010-0044-5} {\bibfield
  {journal} {\bibinfo  {journal} {Annales H. Poincar{\'e}}\ }\textbf {\bibinfo
  {volume} {11}},\ \bibinfo {pages} {351} (\bibinfo {year} {2010})},\ \Eprint
  {https://arxiv.org/abs/0907.4082} {arXiv:0907.4082 [hep-th]} \BibitemShut
  {NoStop}%
\bibitem [{\citenamefont {Taya}\ \emph {et~al.}(2021)\citenamefont {Taya},
  \citenamefont {Fujimori}, \citenamefont {Misumi}, \citenamefont {Nitta},\
  and\ \citenamefont {Sakai}}]{Taya:2020dco}%
  \BibitemOpen
  \bibfield  {author} {\bibinfo {author} {\bibfnamefont {H.}~\bibnamefont
  {Taya}}, \bibinfo {author} {\bibfnamefont {T.}~\bibnamefont {Fujimori}},
  \bibinfo {author} {\bibfnamefont {T.}~\bibnamefont {Misumi}}, \bibinfo
  {author} {\bibfnamefont {M.}~\bibnamefont {Nitta}},\ and\ \bibinfo {author}
  {\bibfnamefont {N.}~\bibnamefont {Sakai}},\ }\href
  {https://doi.org/10.1007/JHEP03(2021)082} {\bibfield  {journal} {\bibinfo
  {journal} {J. High Energy Phys.}\ }\textbf {\bibinfo {volume} {03}},\
  \bibinfo {pages} {082}},\ \Eprint {https://arxiv.org/abs/2010.16080}
  {arXiv:2010.16080 [hep-th]} \BibitemShut {NoStop}%
\bibitem [{\citenamefont {Romatschke}(2017)}]{Romatschke:2017acs}%
  \BibitemOpen
  \bibfield  {author} {\bibinfo {author} {\bibfnamefont {P.}~\bibnamefont
  {Romatschke}},\ }\href {https://doi.org/10.1007/JHEP12(2017)079} {\bibfield
  {journal} {\bibinfo  {journal} {J. High Energy Phys.}\ }\textbf {\bibinfo
  {volume} {12}},\ \bibinfo {pages} {079}},\ \Eprint
  {https://arxiv.org/abs/1710.03234} {arXiv:1710.03234 [hep-th]} \BibitemShut
  {NoStop}%
\bibitem [{\citenamefont {Mitra}\ \emph {et~al.}(2020)\citenamefont {Mitra},
  \citenamefont {Mondkar}, \citenamefont {Mukhopadhyay}, \citenamefont
  {Rebhan},\ and\ \citenamefont {Soloviev}}]{Mitra:2020mei}%
  \BibitemOpen
  \bibfield  {author} {\bibinfo {author} {\bibfnamefont {T.}~\bibnamefont
  {Mitra}}, \bibinfo {author} {\bibfnamefont {S.}~\bibnamefont {Mondkar}},
  \bibinfo {author} {\bibfnamefont {A.}~\bibnamefont {Mukhopadhyay}}, \bibinfo
  {author} {\bibfnamefont {A.}~\bibnamefont {Rebhan}},\ and\ \bibinfo {author}
  {\bibfnamefont {A.}~\bibnamefont {Soloviev}},\ }\href
  {https://doi.org/10.1103/PhysRevResearch.2.043320} {\bibfield  {journal}
  {\bibinfo  {journal} {Phys. Rev. Res.}\ }\textbf {\bibinfo {volume} {2}},\
  \bibinfo {pages} {043320} (\bibinfo {year} {2020})},\ \Eprint
  {https://arxiv.org/abs/2006.09383} {arXiv:2006.09383 [hep-th]} \BibitemShut
  {NoStop}%
\bibitem [{\citenamefont {Blaizot}\ and\ \citenamefont
  {Yan}(2020{\natexlab{a}})}]{Blaizot:2020gql}%
  \BibitemOpen
  \bibfield  {author} {\bibinfo {author} {\bibfnamefont {J.-P.}\ \bibnamefont
  {Blaizot}}\ and\ \bibinfo {author} {\bibfnamefont {L.}~\bibnamefont {Yan}},\
  }\href@noop {} {\  (\bibinfo {year} {2020}{\natexlab{a}})},\ \Eprint
  {https://arxiv.org/abs/2006.08815} {arXiv:2006.08815 [nucl-th]} \BibitemShut
  {NoStop}%
\bibitem [{\citenamefont {Behtash}\ \emph {et~al.}(2020)\citenamefont
  {Behtash}, \citenamefont {Kamata}, \citenamefont {Martinez},\ and\
  \citenamefont {Shi}}]{Behtash:2019qtk}%
  \BibitemOpen
  \bibfield  {author} {\bibinfo {author} {\bibfnamefont {A.}~\bibnamefont
  {Behtash}}, \bibinfo {author} {\bibfnamefont {S.}~\bibnamefont {Kamata}},
  \bibinfo {author} {\bibfnamefont {M.}~\bibnamefont {Martinez}},\ and\
  \bibinfo {author} {\bibfnamefont {H.}~\bibnamefont {Shi}},\ }\href
  {https://doi.org/10.1007/JHEP07(2020)226} {\bibfield  {journal} {\bibinfo
  {journal} {J. High Energy Phys.}\ }\textbf {\bibinfo {volume} {07}},\
  \bibinfo {pages} {226}},\ \Eprint {https://arxiv.org/abs/1911.06406}
  {arXiv:1911.06406 [hep-th]} \BibitemShut {NoStop}%
\bibitem [{\citenamefont {Jaiswal}\ \emph {et~al.}(2019)\citenamefont
  {Jaiswal}, \citenamefont {Chattopadhyay}, \citenamefont {Jaiswal},
  \citenamefont {Pal},\ and\ \citenamefont {Heinz}}]{Jaiswal:2019cju}%
  \BibitemOpen
  \bibfield  {author} {\bibinfo {author} {\bibfnamefont {S.}~\bibnamefont
  {Jaiswal}}, \bibinfo {author} {\bibfnamefont {C.}~\bibnamefont
  {Chattopadhyay}}, \bibinfo {author} {\bibfnamefont {A.}~\bibnamefont
  {Jaiswal}}, \bibinfo {author} {\bibfnamefont {S.}~\bibnamefont {Pal}},\ and\
  \bibinfo {author} {\bibfnamefont {U.}~\bibnamefont {Heinz}},\ }\href
  {https://doi.org/10.1103/PhysRevC.100.034901} {\bibfield  {journal} {\bibinfo
   {journal} {Phys. Rev. C}\ }\textbf {\bibinfo {volume} {100}},\ \bibinfo
  {pages} {034901} (\bibinfo {year} {2019})},\ \Eprint
  {https://arxiv.org/abs/1907.07965} {arXiv:1907.07965 [nucl-th]} \BibitemShut
  {NoStop}%
\bibitem [{\citenamefont {Kurkela}\ \emph {et~al.}(2020)\citenamefont
  {Kurkela}, \citenamefont {van~der Schee}, \citenamefont {Wiedemann},\ and\
  \citenamefont {Wu}}]{Kurkela:2019set}%
  \BibitemOpen
  \bibfield  {author} {\bibinfo {author} {\bibfnamefont {A.}~\bibnamefont
  {Kurkela}}, \bibinfo {author} {\bibfnamefont {W.}~\bibnamefont {van~der
  Schee}}, \bibinfo {author} {\bibfnamefont {U.~A.}\ \bibnamefont
  {Wiedemann}},\ and\ \bibinfo {author} {\bibfnamefont {B.}~\bibnamefont
  {Wu}},\ }\href {https://doi.org/10.1103/PhysRevLett.124.102301} {\bibfield
  {journal} {\bibinfo  {journal} {Phys. Rev. Lett.}\ }\textbf {\bibinfo
  {volume} {124}},\ \bibinfo {pages} {102301} (\bibinfo {year} {2020})},\
  \Eprint {https://arxiv.org/abs/1907.08101} {arXiv:1907.08101 [hep-th]}
  \BibitemShut {NoStop}%
\bibitem [{\citenamefont {Blaizot}\ and\ \citenamefont
  {Yan}(2020{\natexlab{b}})}]{Blaizot:2019scw}%
  \BibitemOpen
  \bibfield  {author} {\bibinfo {author} {\bibfnamefont {J.-P.}\ \bibnamefont
  {Blaizot}}\ and\ \bibinfo {author} {\bibfnamefont {L.}~\bibnamefont {Yan}},\
  }\href {https://doi.org/10.1016/j.aop.2019.167993} {\bibfield  {journal}
  {\bibinfo  {journal} {Annals Phys.}\ }\textbf {\bibinfo {volume} {412}},\
  \bibinfo {pages} {167993} (\bibinfo {year} {2020}{\natexlab{b}})},\ \Eprint
  {https://arxiv.org/abs/1904.08677} {arXiv:1904.08677 [nucl-th]} \BibitemShut
  {NoStop}%
\bibitem [{\citenamefont {Behtash}\ \emph
  {et~al.}(2019{\natexlab{a}})\citenamefont {Behtash}, \citenamefont {Kamata},
  \citenamefont {Martinez},\ and\ \citenamefont {Shi}}]{Behtash:2019txb}%
  \BibitemOpen
  \bibfield  {author} {\bibinfo {author} {\bibfnamefont {A.}~\bibnamefont
  {Behtash}}, \bibinfo {author} {\bibfnamefont {S.}~\bibnamefont {Kamata}},
  \bibinfo {author} {\bibfnamefont {M.}~\bibnamefont {Martinez}},\ and\
  \bibinfo {author} {\bibfnamefont {H.}~\bibnamefont {Shi}},\ }\href
  {https://doi.org/10.1103/PhysRevD.99.116012} {\bibfield  {journal} {\bibinfo
  {journal} {Phys. Rev. D}\ }\textbf {\bibinfo {volume} {99}},\ \bibinfo
  {pages} {116012} (\bibinfo {year} {2019}{\natexlab{a}})},\ \Eprint
  {https://arxiv.org/abs/1901.08632} {arXiv:1901.08632 [hep-th]} \BibitemShut
  {NoStop}%
\bibitem [{\citenamefont {Strickland}(2018)}]{Strickland:2018ayk}%
  \BibitemOpen
  \bibfield  {author} {\bibinfo {author} {\bibfnamefont {M.}~\bibnamefont
  {Strickland}},\ }\href {https://doi.org/10.1007/JHEP12(2018)128} {\bibfield
  {journal} {\bibinfo  {journal} {J. High Energy Phys.}\ }\textbf {\bibinfo
  {volume} {12}},\ \bibinfo {pages} {128}},\ \Eprint
  {https://arxiv.org/abs/1809.01200} {arXiv:1809.01200 [nucl-th]} \BibitemShut
  {NoStop}%
\bibitem [{\citenamefont {Blaizot}\ and\ \citenamefont
  {Yan}(2018)}]{Blaizot:2017ucy}%
  \BibitemOpen
  \bibfield  {author} {\bibinfo {author} {\bibfnamefont {J.-P.}\ \bibnamefont
  {Blaizot}}\ and\ \bibinfo {author} {\bibfnamefont {L.}~\bibnamefont {Yan}},\
  }\href {https://doi.org/10.1016/j.physletb.2018.02.058} {\bibfield  {journal}
  {\bibinfo  {journal} {Phys. Lett. B}\ }\textbf {\bibinfo {volume} {780}},\
  \bibinfo {pages} {283} (\bibinfo {year} {2018})},\ \Eprint
  {https://arxiv.org/abs/1712.03856} {arXiv:1712.03856 [nucl-th]} \BibitemShut
  {NoStop}%
\bibitem [{\citenamefont {Dash}\ and\ \citenamefont
  {Roy}(2020)}]{Dash:2020zqx}%
  \BibitemOpen
  \bibfield  {author} {\bibinfo {author} {\bibfnamefont {A.}~\bibnamefont
  {Dash}}\ and\ \bibinfo {author} {\bibfnamefont {V.}~\bibnamefont {Roy}},\
  }\href {https://doi.org/10.1016/j.physletb.2020.135481} {\bibfield  {journal}
  {\bibinfo  {journal} {Phys. Lett. B}\ }\textbf {\bibinfo {volume} {806}},\
  \bibinfo {pages} {135481} (\bibinfo {year} {2020})},\ \Eprint
  {https://arxiv.org/abs/2001.10756} {arXiv:2001.10756 [nucl-th]} \BibitemShut
  {NoStop}%
\bibitem [{\citenamefont {Aoki}\ \emph {et~al.}(2006)\citenamefont {Aoki},
  \citenamefont {Endr{\H o}di}, \citenamefont {Fodor}, \citenamefont {Katz},\
  and\ \citenamefont {Szab{\'o}}}]{Aoki_2006}%
  \BibitemOpen
  \bibfield  {author} {\bibinfo {author} {\bibfnamefont {Y.}~\bibnamefont
  {Aoki}}, \bibinfo {author} {\bibfnamefont {G.}~\bibnamefont {Endr{\H o}di}},
  \bibinfo {author} {\bibfnamefont {Z.}~\bibnamefont {Fodor}}, \bibinfo
  {author} {\bibfnamefont {S.~D.}\ \bibnamefont {Katz}},\ and\ \bibinfo
  {author} {\bibfnamefont {K.~K.}\ \bibnamefont {Szab{\'o}}},\ }\href
  {https://doi.org/10.1038/nature05120} {\bibfield  {journal} {\bibinfo
  {journal} {Nature}\ }\textbf {\bibinfo {volume} {443}},\ \bibinfo {pages}
  {675} (\bibinfo {year} {2006})},\ \Eprint
  {https://arxiv.org/abs/hep-lat/0611014} {arXiv:hep-lat/0611014} \BibitemShut
  {NoStop}%
\bibitem [{\citenamefont {Buchel}\ \emph {et~al.}(2016)\citenamefont {Buchel},
  \citenamefont {Heller},\ and\ \citenamefont {Noronha}}]{Buchel:2016cbj}%
  \BibitemOpen
  \bibfield  {author} {\bibinfo {author} {\bibfnamefont {A.}~\bibnamefont
  {Buchel}}, \bibinfo {author} {\bibfnamefont {M.~P.}\ \bibnamefont {Heller}},\
  and\ \bibinfo {author} {\bibfnamefont {J.}~\bibnamefont {Noronha}},\ }\href
  {https://doi.org/10.1103/PhysRevD.94.106011} {\bibfield  {journal} {\bibinfo
  {journal} {Phys. Rev. D}\ }\textbf {\bibinfo {volume} {94}},\ \bibinfo
  {pages} {106011} (\bibinfo {year} {2016})}\BibitemShut {NoStop}%
\bibitem [{\citenamefont {Grad}(1963)}]{grad1963}%
  \BibitemOpen
  \bibfield  {author} {\bibinfo {author} {\bibfnamefont {H.}~\bibnamefont
  {Grad}},\ }\href {https://doi.org/10.1063/1.1706716} {\bibfield  {journal}
  {\bibinfo  {journal} {Phys. of Fluids}\ }\textbf {\bibinfo {volume} {6}},\
  \bibinfo {pages} {147} (\bibinfo {year} {1963})}\BibitemShut {NoStop}%
\bibitem [{\citenamefont {Bazow}\ \emph
  {et~al.}(2016{\natexlab{a}})\citenamefont {Bazow}, \citenamefont {Denicol},
  \citenamefont {Heinz}, \citenamefont {Martinez},\ and\ \citenamefont
  {Noronha}}]{Bazow:2016:ASBEES}%
  \BibitemOpen
  \bibfield  {author} {\bibinfo {author} {\bibfnamefont {D.}~\bibnamefont
  {Bazow}}, \bibinfo {author} {\bibfnamefont {G.~S.}\ \bibnamefont {Denicol}},
  \bibinfo {author} {\bibfnamefont {U.}~\bibnamefont {Heinz}}, \bibinfo
  {author} {\bibfnamefont {M.}~\bibnamefont {Martinez}},\ and\ \bibinfo
  {author} {\bibfnamefont {J.}~\bibnamefont {Noronha}},\ }\href
  {https://doi.org/10.1103/PhysRevLett.116.022301} {\bibfield  {journal}
  {\bibinfo  {journal} {Phys. Rev. Lett.}\ }\textbf {\bibinfo {volume} {116}},\
  \bibinfo {pages} {022301} (\bibinfo {year} {2016}{\natexlab{a}})},\ \Eprint
  {https://arxiv.org/abs/1507.07834} {arXiv:1507.07834 [hep-th]} \BibitemShut
  {NoStop}%
\bibitem [{\citenamefont {Bazow}\ \emph
  {et~al.}(2016{\natexlab{b}})\citenamefont {Bazow}, \citenamefont {Denicol},
  \citenamefont {Heinz}, \citenamefont {Martinez},\ and\ \citenamefont
  {Noronha}}]{Bazow:2016:NdrBeFs}%
  \BibitemOpen
  \bibfield  {author} {\bibinfo {author} {\bibfnamefont {D.}~\bibnamefont
  {Bazow}}, \bibinfo {author} {\bibfnamefont {G.~S.}\ \bibnamefont {Denicol}},
  \bibinfo {author} {\bibfnamefont {U.}~\bibnamefont {Heinz}}, \bibinfo
  {author} {\bibfnamefont {M.}~\bibnamefont {Martinez}},\ and\ \bibinfo
  {author} {\bibfnamefont {J.}~\bibnamefont {Noronha}},\ }\href
  {https://doi.org/10.1103/PhysRevD.94.125006} {\bibfield  {journal} {\bibinfo
  {journal} {Phys. Rev. D}\ }\textbf {\bibinfo {volume} {94}},\ \bibinfo
  {pages} {125006} (\bibinfo {year} {2016}{\natexlab{b}})},\ \Eprint
  {https://arxiv.org/abs/1607.05245} {arXiv:1607.05245 [hep-ph]} \BibitemShut
  {NoStop}%
\bibitem [{\citenamefont {Chapman}\ and\ \citenamefont
  {Cowling}(1953)}]{chapman53}%
  \BibitemOpen
  \bibfield  {author} {\bibinfo {author} {\bibfnamefont {S.}~\bibnamefont
  {Chapman}}\ and\ \bibinfo {author} {\bibfnamefont {T.~G.}\ \bibnamefont
  {Cowling}},\ }\href@noop {} {\emph {\bibinfo {title} {The Mathematical Theory
  of Non-Uniform Gases}}}\ (\bibinfo  {publisher} {Cambridge University
  Press},\ \bibinfo {address} {Cambride},\ \bibinfo {year} {1953})\BibitemShut
  {NoStop}%
\bibitem [{\citenamefont {Cercignani}\ and\ \citenamefont
  {Kremer}(2002)}]{CercignaniKremer}%
  \BibitemOpen
  \bibfield  {author} {\bibinfo {author} {\bibfnamefont {C.}~\bibnamefont
  {Cercignani}}\ and\ \bibinfo {author} {\bibfnamefont {G.~M.}\ \bibnamefont
  {Kremer}},\ }\href@noop {} {\emph {\bibinfo {title} {The relativistic
  Boltzmann equation : theory and applications}}}\ (\bibinfo  {publisher}
  {Birkh{\"a}user, Basel},\ \bibinfo {year} {2002})\BibitemShut {NoStop}%
\bibitem [{\citenamefont {Aniceto}\ \emph {et~al.}(2012)\citenamefont
  {Aniceto}, \citenamefont {Schiappa},\ and\ \citenamefont
  {Vonk}}]{Aniceto:2011nu}%
  \BibitemOpen
  \bibfield  {author} {\bibinfo {author} {\bibfnamefont {I.}~\bibnamefont
  {Aniceto}}, \bibinfo {author} {\bibfnamefont {R.}~\bibnamefont {Schiappa}},\
  and\ \bibinfo {author} {\bibfnamefont {M.}~\bibnamefont {Vonk}},\ }\href
  {https://doi.org/10.4310/CNTP.2012.v6.n2.a3} {\bibfield  {journal} {\bibinfo
  {journal} {Commun. Num. Theor. Phys.}\ }\textbf {\bibinfo {volume} {6}},\
  \bibinfo {pages} {339} (\bibinfo {year} {2012})},\ \Eprint
  {https://arxiv.org/abs/1106.5922} {arXiv:1106.5922 [hep-th]} \BibitemShut
  {NoStop}%
\bibitem [{\citenamefont {Cherman}\ \emph {et~al.}(2015)\citenamefont
  {Cherman}, \citenamefont {Koroteev},\ and\ \citenamefont
  {{\"U}nsal}}]{ChermanKoroteevUnsal}%
  \BibitemOpen
  \bibfield  {author} {\bibinfo {author} {\bibfnamefont {A.}~\bibnamefont
  {Cherman}}, \bibinfo {author} {\bibfnamefont {P.}~\bibnamefont {Koroteev}},\
  and\ \bibinfo {author} {\bibfnamefont {M.}~\bibnamefont {{\"U}nsal}},\ }\href
  {https://doi.org/10.1063/1.4921155} {\bibfield  {journal} {\bibinfo
  {journal} {J. Math. Phys.}\ }\textbf {\bibinfo {volume} {56}},\ \bibinfo
  {pages} {053505} (\bibinfo {year} {2015})},\ \Eprint
  {https://arxiv.org/abs/1410.0388} {arXiv:1410.0388 [hep-th]} \BibitemShut
  {NoStop}%
\bibitem [{\citenamefont {Dorigoni}(2014)}]{dorigoni14}%
  \BibitemOpen
  \bibfield  {author} {\bibinfo {author} {\bibfnamefont {D.}~\bibnamefont
  {Dorigoni}},\ }\href {https://doi.org/10.1016/j.aop.2019.167914} {\bibfield
  {journal} {\bibinfo  {journal} {Annals Phys.}\ }\textbf {\bibinfo {volume}
  {409}},\ \bibinfo {pages} {167914} (\bibinfo {year} {2014})},\ \Eprint
  {https://arxiv.org/abs/1411.3585} {arXiv:1411.3585 [hep-th]} \BibitemShut
  {NoStop}%
\bibitem [{\citenamefont {Dunne}\ and\ \citenamefont
  {Unsal}(2014)}]{Dunne:2014bca}%
  \BibitemOpen
  \bibfield  {author} {\bibinfo {author} {\bibfnamefont {G.~V.}\ \bibnamefont
  {Dunne}}\ and\ \bibinfo {author} {\bibfnamefont {M.}~\bibnamefont {Unsal}},\
  }\href {https://doi.org/10.1103/PhysRevD.89.105009} {\bibfield  {journal}
  {\bibinfo  {journal} {Phys. Rev. D}\ }\textbf {\bibinfo {volume} {89}},\
  \bibinfo {pages} {105009} (\bibinfo {year} {2014})},\ \Eprint
  {https://arxiv.org/abs/1401.5202} {arXiv:1401.5202 [hep-th]} \BibitemShut
  {NoStop}%
\bibitem [{\citenamefont {Carroll}(2004)}]{Carroll}%
  \BibitemOpen
  \bibfield  {author} {\bibinfo {author} {\bibfnamefont {S.~M.}\ \bibnamefont
  {Carroll}},\ }\href@noop {} {\emph {\bibinfo {title} {Spacetime and Geometry:
  An Introduction to General Relativity}}}\ (\bibinfo  {publisher} {Addison
  Wesley},\ \bibinfo {address} {San Francisco},\ \bibinfo {year}
  {2004})\BibitemShut {NoStop}%
\bibitem [{\citenamefont {Sch{\"a}fer}(2014)}]{schafer2014}%
  \BibitemOpen
  \bibfield  {author} {\bibinfo {author} {\bibfnamefont {T.}~\bibnamefont
  {Sch{\"a}fer}},\ }\href {https://doi.org/10.1146/annurev-nucl-102313-025439}
  {\bibfield  {journal} {\bibinfo  {journal} {Ann. Rev. Nucl. Part. Sci.}\
  }\textbf {\bibinfo {volume} {64}},\ \bibinfo {pages} {125} (\bibinfo {year}
  {2014})},\ \Eprint {https://arxiv.org/abs/1403.0653} {arXiv:1403.0653
  [hep-th]} \BibitemShut {NoStop}%
\bibitem [{\citenamefont {Brewer}\ and\ \citenamefont
  {Romatschke}(2015)}]{brewer2015}%
  \BibitemOpen
  \bibfield  {author} {\bibinfo {author} {\bibfnamefont {J.}~\bibnamefont
  {Brewer}}\ and\ \bibinfo {author} {\bibfnamefont {P.}~\bibnamefont
  {Romatschke}},\ }\href {https://doi.org/10.1103/PhysRevLett.115.190404}
  {\bibfield  {journal} {\bibinfo  {journal} {Phys. Rev. Lett.}\ }\textbf
  {\bibinfo {volume} {115}},\ \bibinfo {pages} {190404} (\bibinfo {year}
  {2015})},\ \Eprint {https://arxiv.org/abs/1508.01199} {arXiv:1508.01199
  [hep-th]} \BibitemShut {NoStop}%
\bibitem [{\citenamefont {Baier}\ \emph {et~al.}(2008)\citenamefont {Baier},
  \citenamefont {Romatschke}, \citenamefont {Son}, \citenamefont {Starinets},\
  and\ \citenamefont {Stephanov}}]{Baier_2008}%
  \BibitemOpen
  \bibfield  {author} {\bibinfo {author} {\bibfnamefont {R.}~\bibnamefont
  {Baier}}, \bibinfo {author} {\bibfnamefont {P.}~\bibnamefont {Romatschke}},
  \bibinfo {author} {\bibfnamefont {D.~T.}\ \bibnamefont {Son}}, \bibinfo
  {author} {\bibfnamefont {A.~O.}\ \bibnamefont {Starinets}},\ and\ \bibinfo
  {author} {\bibfnamefont {M.~A.}\ \bibnamefont {Stephanov}},\ }\href
  {https://doi.org/10.1088/1126-6708/2008/04/100} {\bibfield  {journal}
  {\bibinfo  {journal} {J. High Energy Phys.}\ }\textbf {\bibinfo {volume}
  {04}},\ \bibinfo {pages} {100}},\ \Eprint {https://arxiv.org/abs/0823.3226}
  {arXiv:0823.3226 [hep-th]} \BibitemShut {NoStop}%
\bibitem [{\citenamefont {Behtash}\ \emph {et~al.}(2021)\citenamefont
  {Behtash}, \citenamefont {Kamata}, \citenamefont {Martinez}, \citenamefont
  {Sch{\"a}fer},\ and\ \citenamefont {Skokov}}]{Behtash:2021coeffRG}%
  \BibitemOpen
  \bibfield  {author} {\bibinfo {author} {\bibfnamefont {A.}~\bibnamefont
  {Behtash}}, \bibinfo {author} {\bibfnamefont {S.}~\bibnamefont {Kamata}},
  \bibinfo {author} {\bibfnamefont {M.}~\bibnamefont {Martinez}}, \bibinfo
  {author} {\bibfnamefont {T.}~\bibnamefont {Sch{\"a}fer}},\ and\ \bibinfo
  {author} {\bibfnamefont {V.}~\bibnamefont {Skokov}},\ }\href
  {https://doi.org/10.1103/PhysRevD.103.056010} {\bibfield  {journal} {\bibinfo
   {journal} {Phys. Rev. D}\ }\textbf {\bibinfo {volume} {103}},\ \bibinfo
  {pages} {056010} (\bibinfo {year} {2021})},\ \Eprint
  {https://arxiv.org/abs/2011.08235} {arXiv:2011.08235 [hep-ph]} \BibitemShut
  {NoStop}%
\bibitem [{\citenamefont {Behtash}\ \emph
  {et~al.}(2019{\natexlab{b}})\citenamefont {Behtash}, \citenamefont
  {Cruz-Camacho}, \citenamefont {Kamata},\ and\ \citenamefont
  {Martinez}}]{Behtash:2018moe}%
  \BibitemOpen
  \bibfield  {author} {\bibinfo {author} {\bibfnamefont {A.}~\bibnamefont
  {Behtash}}, \bibinfo {author} {\bibfnamefont {C.~N.}\ \bibnamefont
  {Cruz-Camacho}}, \bibinfo {author} {\bibfnamefont {S.}~\bibnamefont
  {Kamata}},\ and\ \bibinfo {author} {\bibfnamefont {M.}~\bibnamefont
  {Martinez}},\ }\href {https://doi.org/10.1016/j.physletb.2019.134914}
  {\bibfield  {journal} {\bibinfo  {journal} {Phys. Lett. B}\ }\textbf
  {\bibinfo {volume} {797}},\ \bibinfo {pages} {134914} (\bibinfo {year}
  {2019}{\natexlab{b}})},\ \Eprint {https://arxiv.org/abs/1805.07881}
  {arXiv:1805.07881 [hep-th]} \BibitemShut {NoStop}%
\bibitem [{\citenamefont {Lublinsky}\ and\ \citenamefont
  {Shuryak}(2007)}]{Lublinsky:2007mm}%
  \BibitemOpen
  \bibfield  {author} {\bibinfo {author} {\bibfnamefont {M.}~\bibnamefont
  {Lublinsky}}\ and\ \bibinfo {author} {\bibfnamefont {E.}~\bibnamefont
  {Shuryak}},\ }\href {https://doi.org/10.1103/PhysRevC.76.021901} {\bibfield
  {journal} {\bibinfo  {journal} {Phys. Rev. C}\ }\textbf {\bibinfo {volume}
  {76}},\ \bibinfo {pages} {021901} (\bibinfo {year} {2007})},\ \Eprint
  {https://arxiv.org/abs/0704.1647} {arXiv:0704.1647 [hep-th]} \BibitemShut
  {NoStop}%
\bibitem [{\citenamefont {Bu}\ and\ \citenamefont
  {Lublinsky}(2014)}]{Bu:2014sia}%
  \BibitemOpen
  \bibfield  {author} {\bibinfo {author} {\bibfnamefont {Y.}~\bibnamefont
  {Bu}}\ and\ \bibinfo {author} {\bibfnamefont {M.}~\bibnamefont {Lublinsky}},\
  }\href {https://doi.org/10.1103/PhysRevD.90.086003} {\bibfield  {journal}
  {\bibinfo  {journal} {Phys. Rev. D}\ }\textbf {\bibinfo {volume} {90}},\
  \bibinfo {pages} {086003} (\bibinfo {year} {2014})},\ \Eprint
  {https://arxiv.org/abs/1406.7222} {arXiv:1406.7222 [hep-th]} \BibitemShut
  {NoStop}%
\bibitem [{\citenamefont {McNelis}\ and\ \citenamefont
  {Heinz}(2020)}]{McNelis:2020:Hgrkt}%
  \BibitemOpen
  \bibfield  {author} {\bibinfo {author} {\bibfnamefont {M.}~\bibnamefont
  {McNelis}}\ and\ \bibinfo {author} {\bibfnamefont {U.}~\bibnamefont
  {Heinz}},\ }\href {https://doi.org/10.1103/PhysRevC.101.054901} {\bibfield
  {journal} {\bibinfo  {journal} {Phys. Rev. C}\ }\textbf {\bibinfo {volume}
  {101}},\ \bibinfo {pages} {054901} (\bibinfo {year} {2020})},\ \Eprint
  {https://arxiv.org/abs/2001.09125} {arXiv:2001.09125 [nucl-th]} \BibitemShut
  {NoStop}%
\bibitem [{\citenamefont {Denicol}\ and\ \citenamefont
  {Noronha}(2016)}]{Denicol:2016bjh}%
  \BibitemOpen
  \bibfield  {author} {\bibinfo {author} {\bibfnamefont {G.~S.}\ \bibnamefont
  {Denicol}}\ and\ \bibinfo {author} {\bibfnamefont {J.}~\bibnamefont
  {Noronha}},\ }\href@noop {} {\  (\bibinfo {year} {2016})},\ \Eprint
  {https://arxiv.org/abs/1608.07869} {arXiv:1608.07869 [nucl-th]} \BibitemShut
  {NoStop}%
\bibitem [{\citenamefont {Chen}\ and\ \citenamefont {Sun}(2017)}]{Chen_2017}%
  \BibitemOpen
  \bibfield  {author} {\bibinfo {author} {\bibfnamefont {N.-X.}\ \bibnamefont
  {Chen}}\ and\ \bibinfo {author} {\bibfnamefont {B.-H.}\ \bibnamefont {Sun}},\
  }\href {https://doi.org/10.1088/0256-307x/34/2/020502} {\bibfield  {journal}
  {\bibinfo  {journal} {Chin. Phys. Lett.}\ }\textbf {\bibinfo {volume} {34}},\
  \bibinfo {pages} {020502} (\bibinfo {year} {2017})}\BibitemShut {NoStop}%
\end{thebibliography}%

\end{document}